\documentclass[sn-mathphys-num,hyperref={colorlinks=true, linkcolor=blue, urlcolor=blue, citecolor=blue}]{sn-jnl}

\usepackage{graphicx}
\usepackage{multirow}
\usepackage{amsmath,amssymb,amsfonts}
\usepackage{amsthm}
\usepackage{mathrsfs}
\usepackage[title]{appendix}
\usepackage{xcolor}
\usepackage{textcomp}
\usepackage{manyfoot}
\usepackage{booktabs}
\usepackage{algorithm}
\usepackage{algorithmicx}
\usepackage{algpseudocode}
\usepackage{listings}
\usepackage{placeins}
\usepackage{tabularx}
\usepackage[nameinlink]{cleveref}
\usepackage{subcaption}
\usepackage[per-mode=symbol]{siunitx}
\usepackage{multicol}
\usepackage{url}

\DeclareSIUnit{\mile}{mi}
\DeclareSIUnit{\hour}{hr}
\crefname{figure}{Figure}{Figures}
\Crefname{figure}{Fig.}{Figs.}
\Crefname{equation}{Eq.}{Eqns.}

\theoremstyle{thmstyleone}

\theoremstyle{thmstyletwo}

\theoremstyle{thmstylethree}

\begin{document}

\title[Article Title]{Hybrid Surrogate-Based Multi-Objective Optimization of Graded BCC Lattices}

\author[1]{\fnm{Jaswanth} \sur{Vellore Gurudev}}\email{vgjaswanth@gmail.com}

\author*[2]{\fnm{Ratna Kumar} \sur{Annabattula}}\email{ratna@iitm.ac.in}

\affil[1]{\orgdiv{Department of Metallurgical and Materials Engineering}, \orgname{Indian Institute of Technology Madras}, \orgaddress{\city{Chennai}, \postcode{600036}, \state{Tamil Nadu}, \country{India}}}

\affil*[2]{\orgdiv{Mechanics of Materials Laboratory, Department of Mechanical Engineering}, \orgname{Indian Institute of Technology Madras}, \orgaddress{\city{Chennai}, \postcode{600036}, \state{Tamil Nadu}, \country{India}}}

\abstract{Functionally graded BCC lattice structures hold promise for applications requiring simultaneous impact absorption and thermal dissipation, yet existing optimization frameworks rely on raw geometric parameters that are spatially blind to features such as the orientation of design gradients. In this study, we optimize density-graded BCC lattices for concurrent crashworthiness and heat dissipation via surrogate-based goal programming, initially using raw truss diameter as design variables. The lattice is discretized into three zones to achieve optimal dimensionality, and Pareto-optimal designs were identified that improve specific energy absorption and peak stresses during collisions while also enhancing its Nusselt number and pressure drop under forced convection relative to a non-graded lattice. Key insights of the effect of material distribution along the gradation axis on the performance are discussed in detail. We then introduce Physics-Informed Geometric Operators (PIGOs), which are scalar quantities derived from both the diameter profile and the triangulated surface mesh as candidate surrogate design variables. Pearson correlation analysis reveals that the raw diameter variables remain competitive, with the porosity gradient emerging as the most broadly predictive PIGO, strongly capturing both peak stress and pressure drop. The Nusselt number resists prediction by all variables tested, confirming that heat transfer is irreducibly multidimensional in this configuration due to competing surface-area and flow-blockage effects. The PIGO framework is topology-agnostic and extends naturally beyond the three-zone parameterization to finer gradations and alternative lattice topologies.}

\keywords{Functionally-graded (FG) lattice structures, Multi-physics optimization, Surrogate modeling, Goal programming}

\maketitle

\section{Introduction}\label{section:introduction}

The physical properties of metamaterials (specifically mechanical metamaterials) are often dictated by their intricate internal geometries and high specific surface areas. The ability to control the geometry enables the design of structures with tailored mechanical, thermal, and even multi-functional attributes that are unattainable with conventional monolithic counterparts \citep{fleck2010micro}. Such materials exhibit properties that are not typically found in nature and are often the result of purpose-driven designs, having found applications in aerospace \citep{zhu2018light, ge2024bioinspired, guo2024development, cai2025dual}, automotive \citep{zhang2022advanced, tilley20243d, satpati2025multi, he2025optimized}, biomedical, and thermal management systems \citep{al2025additive, montemurro2022thermal, wang2024thermal, chen2024metamaterials}.

Lattice structures, a class of metamaterials, are gaining prominence for their tunability, especially since the advent of additive manufacturing \citep{veloso2022overview}. While known for crashworthiness, they are increasingly explored for thermal dissipation as traditional heat sinks face limitations from rising power densities \citep{qian2025effect}. Compared to other cellular structures like metal foams or Triply-Periodic Minimal Surfaces (TPMS), lattices offer a compelling alternative due to their high surface area-to-volume ratio and interconnected pores, which balance efficient thermal conduction with low fluid flow resistance \citep{wadley2007thermal, kim2004convective}. Studies show lattices can be more efficient heat sinks than conventional finned designs \citep{kim2004convective, ejlali2009application, qian2023study}. There exist several attempts in the literature to optimize specific topologies, such as BCC lattices, which show superior heat transfer \citep{shahrzadi2022heat}, and functionally graded structures, which vary geometry to further enhance thermal performance \citep{veloso2022overview, vaissier2019parametric, piacquadio2022experimental, qian2025effect}.

Beyond thermal management, lattice structures possess remarkable mechanical properties for energy absorption and crashworthiness. Their performance is characterized by their deformation mechanism: bending-dominated or stretching-dominated \citep{deshpande2001foam}. Stretching-dominated structures (such as Octet lattices) exhibit high nodal connectivity, making them weight-efficient and offering superior stiffness and strength for load-bearing applications. Conversely, bending-dominated structures (like BCC lattices) have lower connectivity, offering less resistance to bending, which is ideal for energy absorption \citep{zhang2024superior, wang2024design}. The capacity of lattices to endure large compressive strains at a near-constant stress makes them suitable for protective applications \citep{gorguluarslan2022multi, nasrullah2020design}. Researchers are maximizing specific energy absorption by exploring configurations such as BCC and Octet lattices, applying functional grading, developing novel hybrid structures (e.g., FCC-BCC), and using topology optimization to achieve uniform performance regardless of impact direction \citep{murugan2024design, nasrullah2020design, al2018mechanical, kappe2024multi, gorguluarslan2022multi, rahimi2025design, alkhatib2023isotropic}.

Due to the numerous possibilities in lattice cell topologies, dimensions, and spatial arrangements, it is often not feasible to conduct experiments or even numerical simulations, given the associated costs and time. To address this issue, significant efforts have been made to utilize surrogate modeling as a powerful technique to interpolate between discrete simulation/experimental data points, especially for improving the crashworthiness of cellular structures \citep{yin2021design,  costa2022multi, xiao2024optimization, tan2021crashworthiness, wang2025poisson, yang2021crashworthiness}. However, the potential to tune lattice structures to meet multiple functional requirements simultaneously is often overlooked and remains underexplored. While a particular lattice may not perform the best across all considered functional needs, an optimal design that performs reasonably well with minimal trade-offs can still be achieved. Combining surrogate modeling with multi-objective optimization has been shown to yield such optimal structures \citep{costas2014multi, xu2016crash}.

\begin{figure*}[t]
    \centering
    \includegraphics[width=0.9\linewidth]{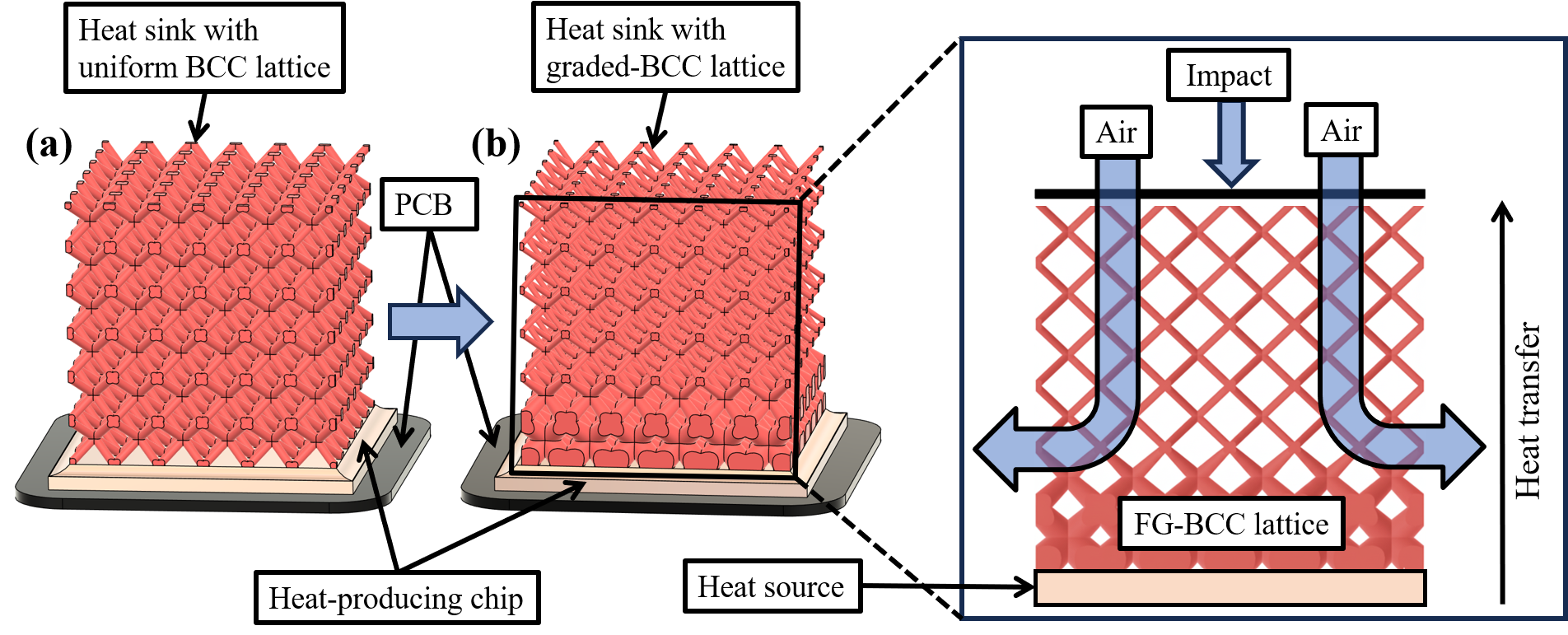}
    \caption{A BCC lattice-based heat sink consisting of (a) uniform struts is replaced with (b) a lattice consisting of diameter-graded struts. A cross-section of the hypothesized FG-BCC lattice with a density gradient is shown. The thinner struts at the impact end reduce both the peak initial stress during impact and the associated pressure drop in the airflow. Meanwhile, thicker struts at the heat source-lattice interface increase the heat flow into the lattice}
    \label{fig:hypothesis}
\end{figure*}

The aim of this paper is to explore BCC lattice structures for their simultaneous energy absorption and heat dissipation capabilities, leveraging the inherent advantages of the BCC topology and functional grading. Such structures can be particularly useful in front-end automotive electronic enclosures for Electronic Control Units (ECUs) due to the presence of strong airflow. Heat sinks made using these lattices for such enclosures would prevent severe component damage upon impact, whilst providing superior passive heat extraction.

When referring to the thermal (mechanical) aspect of the lattice, we denote the end of the lattice in contact with the chip as the heat side (support end) and the other end facing the cold incoming air as the air side (impact end). The key idea behind this study is to create an optimized Functionally-Graded BCC (FG-BCC) lattice with thicker struts on the hot side and thinner struts on the cold side. This is beneficial when compared to a BCC lattice with uniform strut diameter in the following ways:

\begin{itemize}
    \item Thinner struts at the cold side would result in a lower flow resistance during forced convection and would encourage flow mixing within the porous structure of the lattice.
    \item Thicker struts at the heat side directly translate to improved heat extraction from the heat source (governed by Fourier's law of heat conduction).
\end{itemize}

This hypothesis, as illustrated in \Cref{fig:hypothesis}, drives this study to identify the optimal design that simultaneously maximizes both thermal dissipation and impact performance, and to examine the trade-offs that arise in this scenario. We define the vertical three-zone discretization of an FG-BCC lattice, describe the simulation frameworks used in this study, formulate the optimization problem, and build surrogates to identify the Pareto-optimal design.

We start with a periodic BCC lattice consisting of struts with a uniform diameter of \SI{0.78}{\mm} (relative density, $\rho$ = 0.17). This uniform BCC "ground" structure is used as a Representative Volume Element (RVE) for baseline comparison, a configuration validated by \cite{pasvanti2019lattice} and \cite{wu2020modeling} for capturing effective lattice properties. This structure provides a consistent benchmark to quantify the performance gains of the proposed functionally graded designs. Numerical simulations of the ground structure were conducted to establish baseline metrics for heat dissipation and energy absorption. Abaqus$^{\textregistered}$ Explicit \citep{AbaqusExplicit2023} was used to perform finite element simulations for the impact study. Thermal dissipation performance was evaluated by performing Computational Fluid Dynamics (CFD) simulations using the Ansys Fluent 2025 R1 software \citep{AnsysFluent2025R1}. Due to the inherent nature of the problem, trade-offs exist between thermal performance and impact performance, which are discussed in detail. We propose a gradation strategy, as explained in \Cref{subsec:geometry_description}, to explore various relative density gradients across the lattice's height and determine the optimal lattice design. Multi-objective optimization, aided by surrogate modeling, is employed to simultaneously maximize thermal performance and impact performance by exploring the design space. 

This study also examines the effectiveness of using raw truss diameters as primary design variables using Pearson correlation analysis. While effective within a given discretized design space, this raw representation has three fundamental limitations:

\begin{itemize}
    \item Spatial blindness: Raw diameter values encode only endpoint magnitudes, not the shape of the gradient profile. Two lattices with the same diameter values but different intermediate geometries would be indistinguishable.
    \item Topology lock-in: The surrogate is trained on a specific three-zone discretization. Extension to finer or continuous gradations requires complete retraining.
    \item Physical opacity: The surrogate here is forced to learn the physics of bending-dominated collapse and forced convection solely from raw geometric coordinates.
\end{itemize}

We then propose using various scalar quantities, known as Physics-Informed Geometric Operators (PIGOs), as proxies for our performance metrics. The surrogates would then operate in a space where input variables are directly connected to the underlying physics (for example, slenderness governs buckling and porosity gradient governs flow resistance). We evaluate the predictive power of PIGOs using correlations and recommend a set of surrogate input variables found best for this problem.

\section{Methodology}
To systematically navigate the vast design space and identify optimal lattice configurations, this section formalizes the design variables, describes the simulation setups for both thermal and mechanical analyses, and presents the multi-objective optimization procedure.
\subsection{Lattice Design Description}
\label{subsec:geometry_description}
\begin{figure}[htbp]
    \centering
    \includegraphics[width=0.5\linewidth]{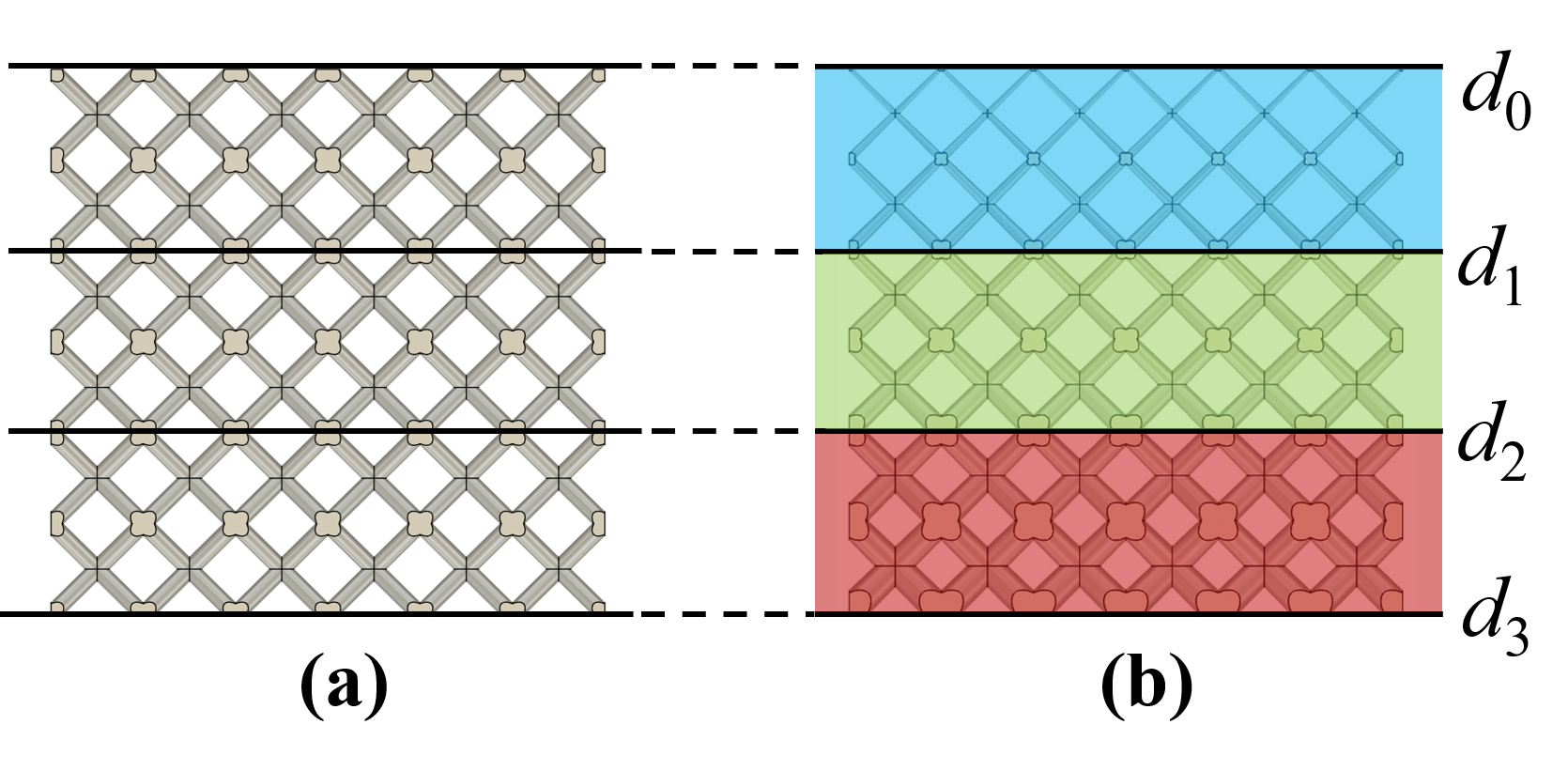}
    \caption{The uniform ground lattice (a) is geometrically compared to a discretized FG-BCC lattice (b), consisting of three planar zones. This discretization yields four distinct diameter values: $d_0$, $d_1$, $d_2$, and $d_3$. These four diameters dictate the geometry of any FG-BCC lattice in this work and will serve as the design variables}
    \label{fig:planar_discretization}
\end{figure}

The mechanical and thermal behavior of lattice structures depends heavily on the nature and degree of gradation of their geometrical features. Here, we vary the relative density of the lattice structure by varying the strut diameter from bottom to top. The FG-BCC lattice structure, consisting of 6 layers (each layer consisting of a $6 \times 6$ arrangement of FG-BCC unit cells), is divided into three planar zones to effectively explore the design space as shown in \Cref{fig:planar_discretization}. As a result, the rate at which the diameter changes is not constant throughout the lattice but instead different for all three zones, resulting in four unique diameter values - $d_0$, $d_1$, $d_2$, and $d_3$, with which the whole lattice structure can be constructed for a given cell size, \SI{4}{\mm} here. Such a setup would allow us to indirectly explore a wide variety of diameter gradients.
To reduce the number of design variables in the problem, we assume $d_0$ = \SI{0.5}{\mm}, owing to the resolution limit of Additive Manufacturing (AM) techniques \citep{chin2020powder, moylan2014additive}. Furthermore, by maintaining the relative density of the FG-BCC lattice the same as the base structure (= 0.17), $d_3$ can be written as a function of $d_1$ and $d_2$, thereby reducing the number of design variables to two. Such a two-dimensional (2D) design space will also enable 3D visualization of the polynomial response surface.

\begin{figure*}[t]
    \centering
    \includegraphics[width=\linewidth]{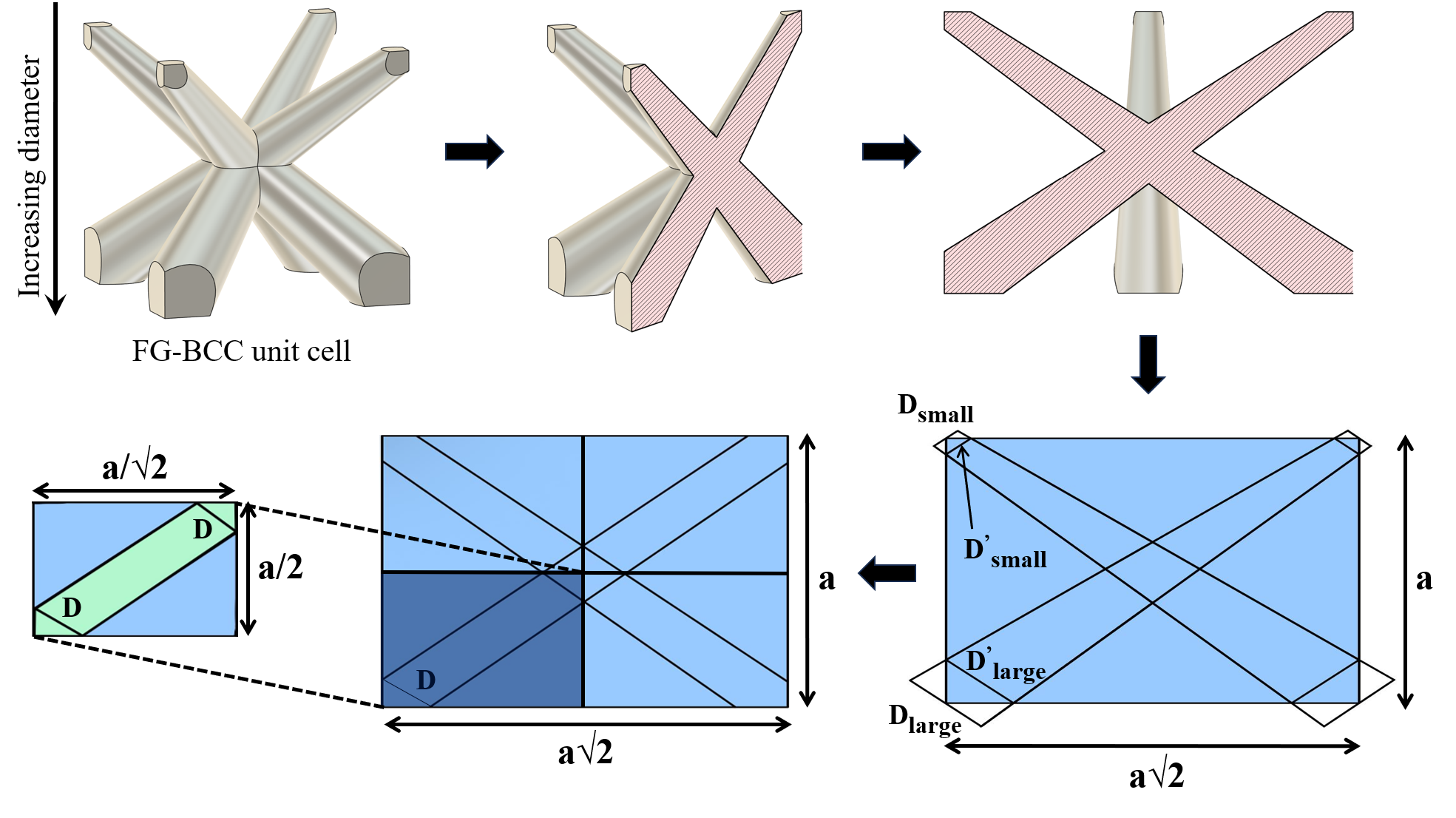}
    \caption{Schematic showing the geometric features of an FG-BCC unit cell. The steps shown in the figure outline the simplified process for estimating area porosity, an indirect, approximate measure of relative density. Area porosity is used as a constraint to generate new lattices, as deriving the analytical expression for the volumetric relative density is complex.}
    \label{fig:rel_density_formulation}
\end{figure*}

An analytical expression for calculating the relative density of FG-BCC lattices is crucial for performing the final step of dimensionality reduction. The relative density of FG-BCC unit cells is generally calculated using the volume fraction of the unit cell. Unfortunately, deriving analytical expressions for calculating the volume occupied by the graded struts is cumbersome due to complex overlap regions at joints and edges. To overcome this problem, we constrain the area fraction rather than the volume fraction in the unit cell. This is done by sectioning the FG-BCC unit cell along the plane with the maximum area fraction, and the analytical expression for the area porosity is calculated as schematically shown in \Cref{fig:rel_density_formulation}.

An FG-BCC unit cell is first cut along its diagonal plane to reveal the smoothly varying strut diameter. We define the large and small diameters at the bottom and top ends of the unit cells as $D_\text{large}$ and $D_\text{small}$, respectively. Our aim is to project this cross-section onto a 2D plane and calculate the area porosity. To skip the calculation of complex area overlaps and simplify the calculation of the cross-sectional area of the unit cell, we construct an `equivalent' cell with a uniform diameter, $D$, that occupies the same cross-sectional area as that of the FG-BCC cell. $D_\text{large}$ and $D_\text{small}$ represent the diameter at the thicker and thinner ends of a strut that intersect the opposite corners of the rectangle as shown in \Cref{fig:rel_density_formulation}. The final expression for $D$ (derivation given in \textit{Supplemental Material}, Section S1) is found to be:

\begin{equation}
    \label{eq:D_eff_calculation}
    \begin{split}
        D &= \dfrac{1}{D_\text{small} - D_\text{large}}\left(\dfrac{D_\text{small}}{2 + \dfrac{2 \sqrt{2}}{3 \sqrt{3} a}} + \dfrac{D_\text{large}}{2 - \dfrac{2 \sqrt{2}}{3 \sqrt{3} a}}\right).
    \end{split}
\end{equation}
The area porosity of the equivalent unit cell is now easily calculated, as explained in \textit{Supplemental Material}, Section S2. The resulting expression for the area porosity of a unit cell (with a uniform strut of diameter $D$) or, equivalently, our FG-BCC unit cell (whose diameter varies from $D_\text{large}$ to $D_\text{small}$) is found to be
\begin{equation}
    \label{eq:area_porosity_calculation}
    \text{Area Porosity,} \;\; \varphi(D) = \Bigg(1-\frac{2D}{a\sqrt{3}}\Bigg)^2.
\end{equation}
This expression for area porosity will now be used to constrain the relative density for generating new FG-BCC lattice designs. We demonstrate that using this approximate method, new FG-BCC lattices were successfully generated with an error of less than 6\% from the target porosity, as shown in \Cref{tab:geometry_info}.
\subsection{Sampling Data Points using Latin Hypercube Sampling (LHS)}
Our objective is to identify the set of $d_1$ and $d_2$ that yields the best overall thermal and impact performance. Due to the high computational cost of running a large number of simulations to cover the entire design space, we employ surrogate modeling to approximate optimal diameter values. Out of the multitude of options, we chose the polynomial response surface method, Thin Plate Spline - Radial Basis Functions (TPS \textendash RBF), due to its overall simplicity and effective interpolation. To achieve a reasonably accurate fit for the surrogates, we select 16 random sample points: 4 to represent the corners of the design space and the remaining 12 uniformly within its bounds. The lower bound of $d_1$, $d_2$, and $d_3$ is set to $d_\text{min}$ = \SI{0.5}{\mm}, the same as that of $d_0$ due to manufacturing constraints. The upper bound is identified by setting three out of the four diameter values to \SI{0.5}{mm} and computing the fourth diameter value subject to the relative density constraint to represent the maximum allowable diameter. Upon calculation using \Cref{eq:D_eff_calculation} and \Cref{eq:area_porosity_calculation}, we get $d_\text{max}$ = \SI{2.58}{mm} as our upper bound for all $d_1$, $d_2$ and $d_3$. We also set $d_0$ $<$ $d_1$ $<$ $d_2$ $<$ $d_3$ to serve our hypothesis of having thicker struts at the base. This naturally results in $d_1 \in [d_0,d_2]$, $d_2\in [d_1, d_3]$ and $d_3\in [d_2, 2.58]$.

The detailed procedure for selecting the final set of 16 points, as shown in \Cref{fig:final_DoE_points}, is described in \textit{Supplemental Material}, Section S3. The final list of the 16 selected points showing the $d_0$, $d_1$, $d_2$, and $d_3$ values along with the deviation in the porosity from the ground structure (resulting from our approximation) is shown in \Cref{tab:geometry_info}. We can now proceed to design CAD models for all 16 points ($d_1$, $d_2$) and then run both impact and forced convection simulations, which will serve as training data for the surrogate model.
\begin{figure}[htbp]
    \centering
    \includegraphics[width=0.5\linewidth]{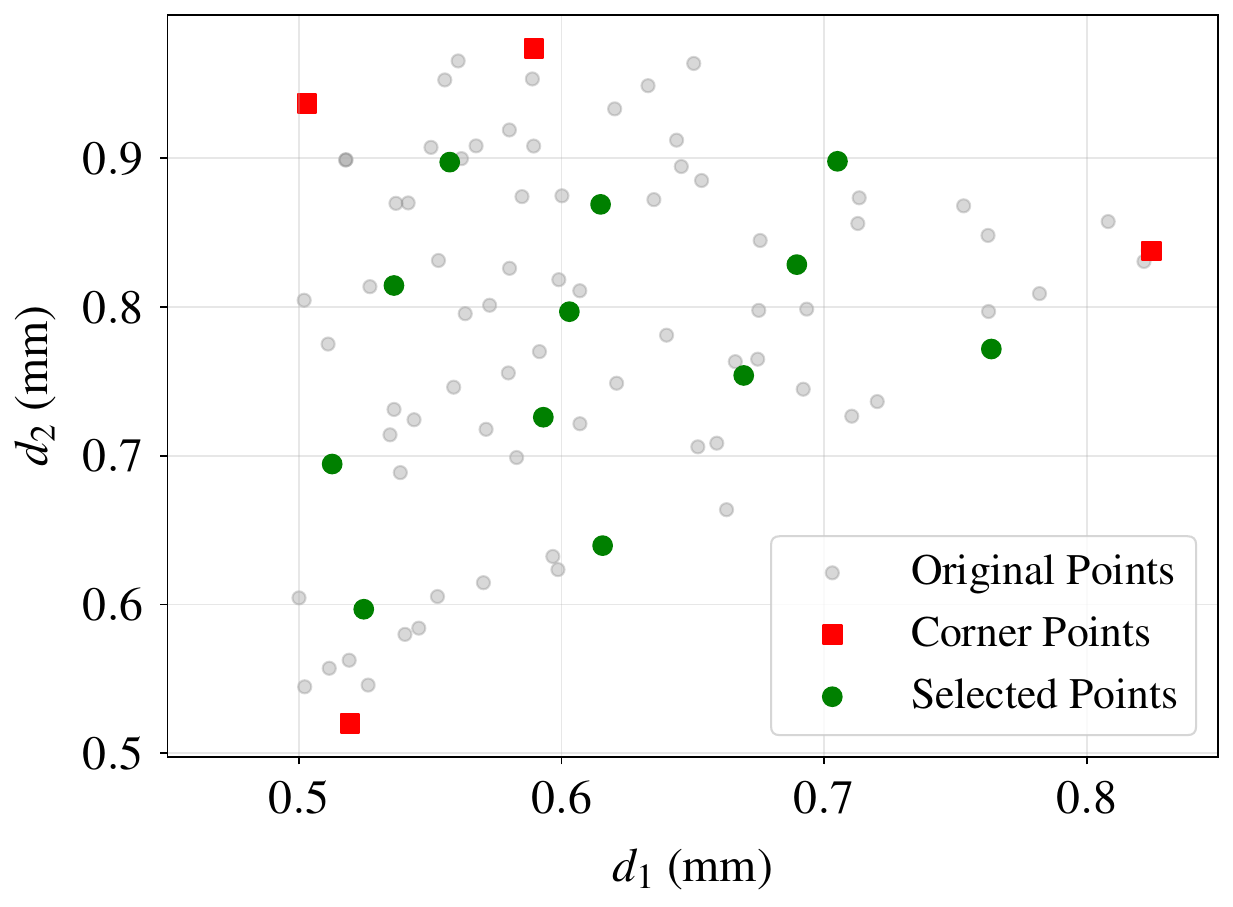}
    \caption{The final set of 16 evenly spaced ($d_1$, $d_2$) points (shown in green) in our feasible region represents our training geometries}
    \label{fig:final_DoE_points}
\end{figure}
\subsection{Forced Convection Simulation}
Ansys Fluent 2025, a commercial software, was used to perform three-dimensional numerical simulations of forced convection in a jet impingement setup. A thin copper chip serves as the heat source, and the lattice to be tested is placed on top of it, and the interface between them is thermally coupled to ensure zero thermal resistance. Fluid (here, air) is forced through a square inlet directly from the top of the lattice along the -ve y direction, and the outlet is modeled to be all around the lattice. The outlet faces are maintained at \SI{0}{\pascal}. Only a quarter of the whole domain is considered owing to the symmetrical nature of the flow. The adjacent faces in contact with both the chip and lattice are set to be symmetric surfaces. The chip is assumed to be made of copper, and the lattice is assumed to be made of Al-6061-T6 alloy. The detailed material properties for both the forced convection and impact simulations are listed in \Cref{tab:material_info}. 

\begin{figure*}[t]
    \centering
    \includegraphics[width=0.7\linewidth]{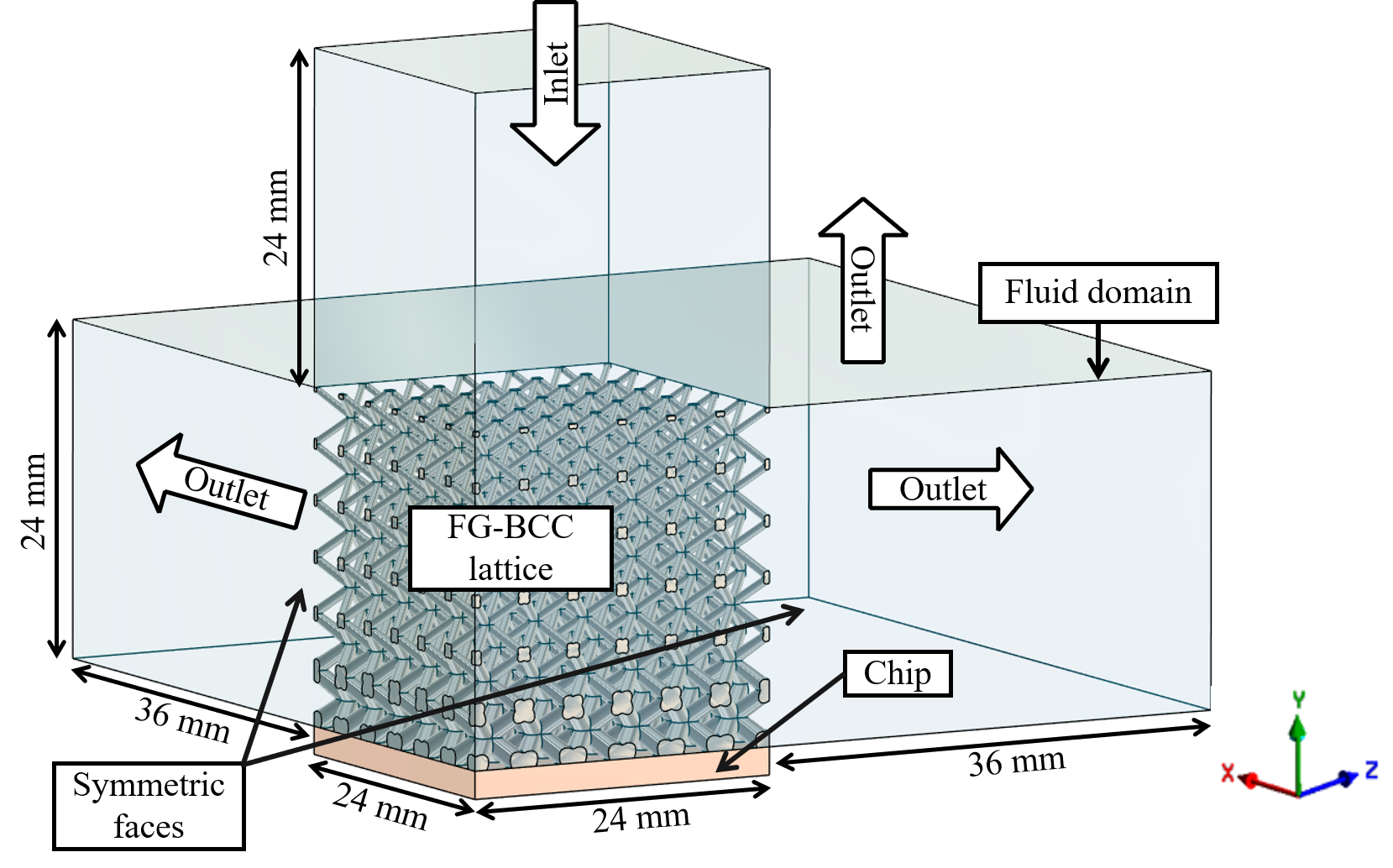}
    \caption{The forced convection simulation setup in Ansys Fluent comprising the FG-BCC-16 lattice}
    \label{fig:ansys_setup}
\end{figure*}

The inlet is positioned 24 mm above the lattice, and the outlets are \SI{36}{\mm} apart in the x and z directions to allow sufficient flow development and prevent reverse flow. Air at \SI{303}{\kelvin} (\SI{30}{\celsius}) is forced from the top at a velocity of \SI{15}{\meter\per\second}. The geometric setup for the forced convection simulation is shown in \Cref{fig:ansys_setup}. We consider the chip to be an infinite plate of negligible thickness that generates a uniform, constant heat flux of roughly \SI[group-separator = {,}]{86806}{\watt\per\meter\squared}. Additionally, due to relatively low fluid velocity \SI{15}{\meter\per\second} (or Mach 0.045) and as the local velocities in the lattice are well below the Mach threshold of 0.3 or roughly \SI{100}{\meter\per\second} as shown in \Cref{fig:fluent_velocity contours}, for all practical purposes, air is assumed to be incompressible \cite{Okawa2021Fundamentals}, and as a result, no density variations exist ($\rho$ is a constant). It is also assumed that the thermophysical properties of air are independent of temperature. Radiative heat transfer and buoyancy effects, which would result in natural convection due to density fluctuations, are also neglected. Simulations are performed under steady-state conditions.

\begin{table*}[t]
    \centering
    \caption{Material properties of Al-6061, Cu and Air}
    \label{tab:material_info}
    \begin{tabular}{ccccccc}
        \toprule
        Material & \begin{tabular}[c]{@{}c@{}}Specific \\ heat capacity \\ (J/kg·K)\end{tabular} & \begin{tabular}[c]{@{}c@{}}Thermal \\ conductivity \\ (W/m·K)\end{tabular} & \begin{tabular}[c]{@{}c@{}}Density \\ (kg/m$^3$)\end{tabular} & \begin{tabular}[c]{@{}c@{}}Young's \\ Modulus \\ (GPa)\end{tabular} & \begin{tabular}[c]{@{}c@{}}Yield \\ Strength \\ (MPa)\end{tabular} & \begin{tabular}[c]{@{}c@{}}Poisson's \\ Ratio\end{tabular} \\ 
        \midrule
        Al-6061 & 871  & 202.40 & 2719  & 70  & 221 & 0.32 \\ 
        Cu         & 381  & 387.60 & 8978  & 117 & 70  & 0.33 \\ 
        Air        & 1006 & 0.0242 & 1.225 & --  & --  & --   \\ 
        \bottomrule
    \end{tabular}
\end{table*}
The SIMPLEC scheme was used to solve the pressure-velocity-coupled equations, and the second-order upwind scheme was employed in high-gradient boundary regions. The simulations were run till 1000 iterations or till the value of the continuity equation reached $10^{-4}$. The convergence of the chip's bulk temperature and the pressure drop across the lattice was also used to decide the termination of the simulation.
The SST (Shear Stress Transport) k-omega turbulence model is a highly versatile and widely used model in ANSYS Fluent for a broad range of CFD simulations, including this study. It effectively combines the best attributes of both the standard k-omega and k-epsilon models. The SST k-omega model excels at predicting flow behaviour in the near-wall region, making it particularly well-suited for applications where boundary-layer effects, flow separation, and adverse pressure gradients are significant, such as in aerodynamics and turbomachinery. By transitioning to a k-epsilon formulation in the free-stream, it avoids some of the limitations of the standard k-$\omega$ model, offering a robust and accurate solution for complex industrial flows without being overly computationally expensive. A detailed review of the governing equations used here is described in \textit{Supplemental Material}, Section S4.

Forced convection setups, especially in our case, involve an air pump at the top of the lattice to push air through its pores. The power required to push air through the lattice is directly proportional to the cube of the pressure drop across the lattice, i.e., $\text{Power}\propto \; (\Delta \text{P})^3$, as observed in many experiments involving simpler setups. A lower power requirement is, hence, tied to a lower pressure drop across the lattice, defined as the average pressure difference between the inlet and the outlet. Due to the outlet being connected to the atmosphere ($P_\text{out} = 0$ Pa),
\begin{equation}
    \Delta {P} = P_\text{in} - P_\text{out} = P_\text{in},
\end{equation}
where $P_\text{in}$ is the area-weighted averaged pressure at the fluid inlet of the setup and $P_\text{out}$ is the area-weighted averaged pressure at the fluid outlet of the setup. Another measure of flow resistance is the unit-less coefficient of friction, $C_\text{f}$, given by
\begin{equation}
    C_\text{f} = \frac{2\Delta \text{P}}{\rho U_\text{in}^2},
\end{equation}
where, $\rho$ is the fluid density and $U_\text{in}$ is the fluid velocity at the inlet.
The Nusselt number is given by,
\begin{equation}
    \text{Nu} = \frac{hL}{k_\text{f}},
\end{equation}
where $L$ is the characteristic length, $k_\text{f}$ is the thermal conductivity of the fluid, and $h$ is the overall heat transfer coefficient along the solid-fluid boundary given by,
\begin{equation}
    h = \frac{Q}{A(T_\text{out} - T_\text{in})} = \frac{q}{T_\text{out} - T_\text{in}},
\end{equation}
where $q$ is the heat flux across the solid-fluid boundary. We can also define an effective thermal performance factor, $\beta$, to take into account both the heat transfer and the power requirement aspects as used in \cite{qian2025effect}, given as
\begin{equation}
    \beta = \frac{\text{Nu}}{C_\text{f}^{1/3}},
\end{equation}

The number of elements and the quality of the mesh in the simulation model can significantly impact the accuracy of the results. A mesh convergence study was conducted to determine a mesh size that balances computational cost and numerical accuracy. It was found that beyond 5.5 million elements, the variables of interest do not increase drastically (shown in \textit{Supplemental Material}, Section S5). Hence, we proceed with using 5.5 million cells (corresponding to an element size of \SI{0.225}{\milli\meter}) for the forced FEA convection simulations.
\subsection{Impact Simulation}
\begin{figure}[!t]
    \centering
    \includegraphics[width=0.5\linewidth]{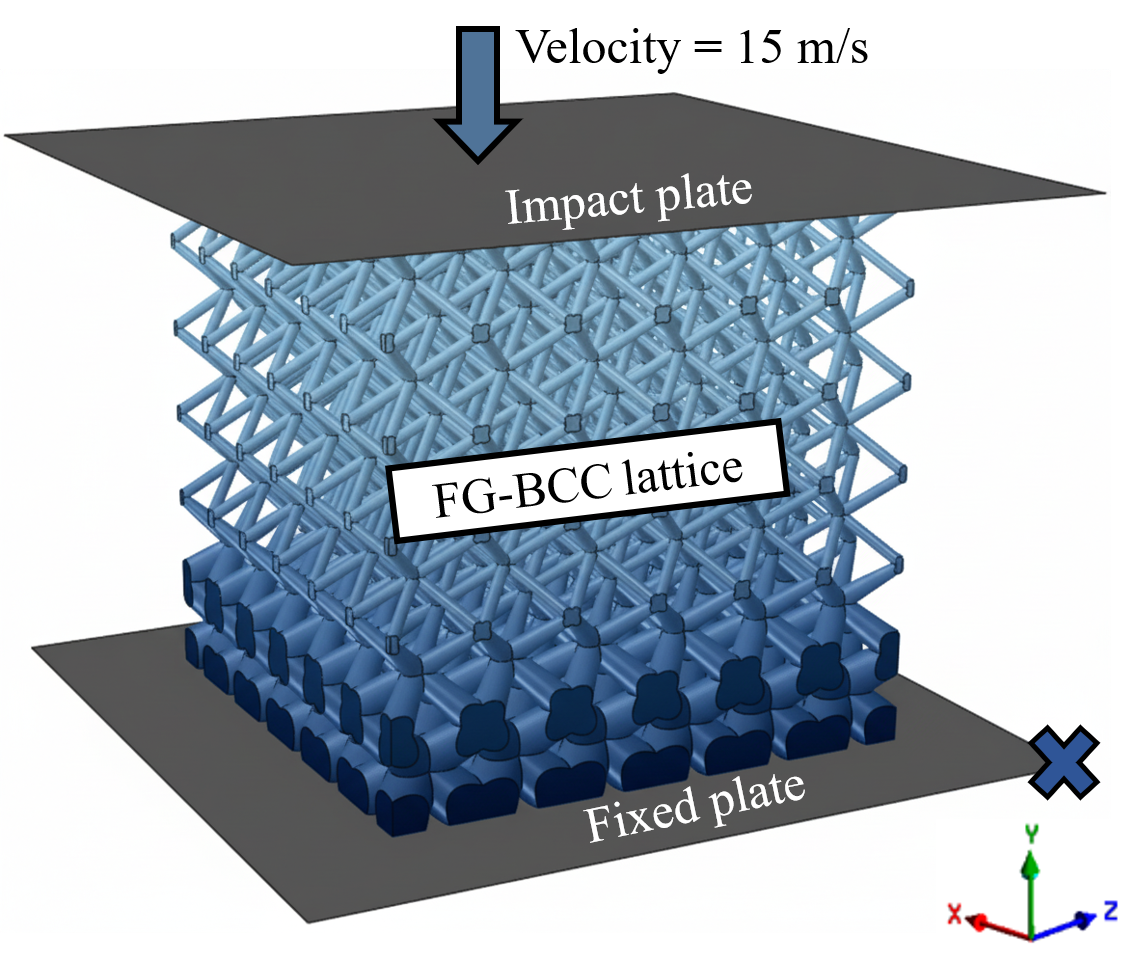}
    \caption{Abaqus/Explicit impact simulation setup of an FG-BCC lattice (FG-BCC-3, specifically), color graded from dark blue at the larger diameter end to light blue at the smaller diameter end, between the fixed and impact plates}
    \label{fig:latticeSetup}
\end{figure}

Dynamic crushing simulations were carried out using a commercial finite element software, Abaqus/Explicit, to study the energy absorption of the FG-BCC lattices. As illustrated in \Cref{fig:latticeSetup}, the simulation consists of the FG-BCC lattice of interest placed between two rigid plates, one plate fixed at the bottom and the other plate moving towards the fixed plate at a constant velocity of \SI{15}{\meter\per\second}, crushing the FG-BCC lattice placed in between. Most car accidents occur below \SI{40}{\mile\per\hour} (\SI{18}{\meter\per\second}) \cite{impact_medical_2023}, and \SI{55}{\kilo\meter\per\hour} (\SI{15.3}{\meter\per\second}) has been concluded as the critical collision velocity by lawyers with respect to injuries \cite{justicepays_sarasota_speed_2025}. For these reasons and to ensure similar scenarios for both impact absorption and thermal dissipation, we set \SI{15}{\meter\per\second} as the impact velocity in this study.

General surface-to-surface contacts are defined in the simulation setup. Tangential surface interactions are assumed to be frictional, with static and dynamic friction coefficients of 0.2 and 0.15, respectively. Normal surface interactions are assumed to be general hard contacts. The impact plate is restricted to move only in the direction of impact, while the fixed plate remains stationary, with no displacement allowed.

Similar to the forced convection simulations, a mesh-convergence study to determine the optimal mesh size that balances accuracy and computational cost. Minimal accuracy improvement is observed beyond a mesh size of 0.25 mm (as shown in \textit{Supplemental Material}, Section S6); therefore, we use this size for discretization.

Al-6061-T6 (see \Cref{tab:material_info}) is chosen as the material of choice due to its relatively high strength-to-weight ratio, thermal conductivity, and corrosion resistance, especially in heat exchangers. A linear-elastic, perfectly plastic material model is assumed for simplicity, and Strain hardening and strain-rate effects are neglected.

\subsubsection{Metrics of Interest for Impact Problem}
\begin{figure}[htbp]
    \centering
    \includegraphics[width=0.5\linewidth]{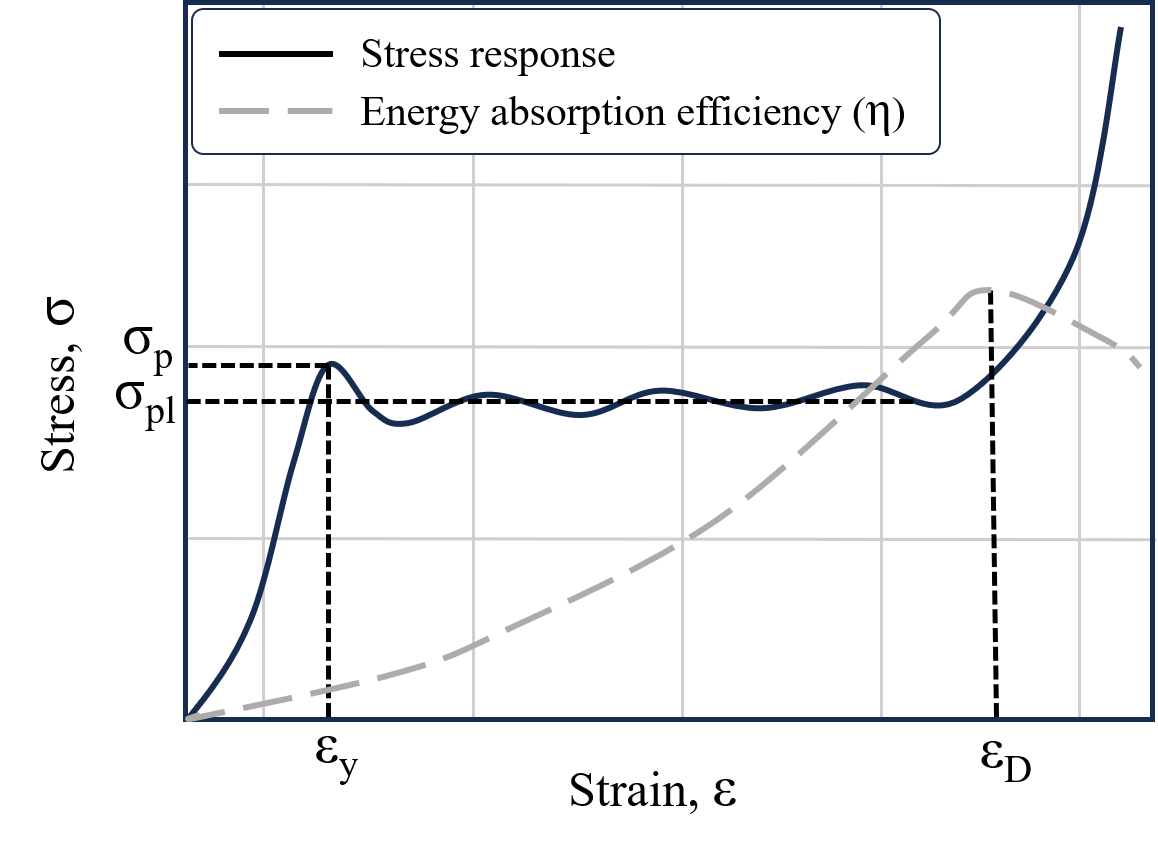}
    \caption{A sample stress-strain curve along with its energy absorption efficiency ($\eta$) curve of a general cellular material under compression}\label{fig:samplestressstrain}
\end{figure}

Load-versus-displacement data, the most common way to represent the compressive behavior of cellular structures, is extracted from impact simulations for post-processing. The compressive load experienced by the fixed plate is divided by the initial cross-sectional area of the lattice ($\SI{24}{\mm} \times \SI{24}{\mm}$) to calculate compressive stress, $\sigma$. Compressive strain, $\epsilon$, is calculated by dividing the displacement by the initial lattice length. The stress–strain response of cellular structures under dynamic compression typically consists of three regions: the initial elastic compression; the plastic deformation and collapse of the cellular structure at approximately a steady stress; and, finally, the densification region, where the structure begins to behave similarly to the bulk material. A representative stress-strain curve showing these three regions is shown in \Cref{fig:samplestressstrain}. 

The energy absorbed by the lattice, represented by the Plastic Dissipation Energy (PDE), is extracted from the Abaqus simulation results. Specific Energy Absorption (SEA), defined here as the plastic energy absorbed per unit mass of the lattice (see \Cref{eq:SEA}), is a widely used metric for comparing lattice structures based on their impact-absorption capabilities. SEA is defined as:
\begin{equation}
    \text{SEA}=\frac{\text{PDE}}{\text{Mass of the lattice}}\label{eq:SEA}.
\end{equation}
The peak stress, $\sigma_\text{p}$, that the cellular structures experience before the onset of plastic deformation is another important metric of study (see \Cref{fig:samplestressstrain}). It represents the maximum stress experienced by the fixed plate (or, in a practical case, the passengers in a vehicle) before or during the initial yielding of a few struts in the lattice. Although the stresses at the fixed end might cross $\sigma_\text{p}$ during the later stages of plastic deformation, the initial  $\sigma_\text{p}$ is also associated with the highest amount of jerk (rate of change of acceleration), due to the sharp rise in stress values during the initial stages. High levels of jerk are hazardous, especially for passenger safety. Hence, $\sigma_\text{p}$ is a crucial metric that is examined in depth in this study, and we aim to minimize it.
The onset strain of densification, $\varepsilon_\text{D}$, marks the beginning of cell wall interaction (see \Cref{fig:samplestressstrain}). At this point, the cell walls begin to contact, suppressing the deformation mechanisms that characterize the plateau region of the stress-strain curve and enhancing the material's compressive resistance. We do not use the term' densification strain,' as it refers to a more advanced stage in which the cellular material is fully compacted. At this strain, the cell walls are jammed together, leading to a steep increase in the slope of the stress-strain curve. Beyond the densification strain, the compressive properties of the cellular solid are dominated by the properties of the solid material from which the cell walls are made. Hence, the onset strain of densification is another crucial parameter for evaluating the impact absorption performance of cellular structures \cite{li2006compressive}. Various methods to find $\varepsilon_\text{D}$ are defined in the existing literature. Out of these, we use the definition given by \citet{tan2002inertia}, which defines $\varepsilon_\text{D}$ as the stationary point at which the Energy Absorption Efficiency ($\eta$) reaches a global maximum. The energy absorption efficiency ($\eta$) is defined as the energy absorbed up to a given nominal strain, $\epsilon_\text{a}$, normalized by the corresponding stress value as follows:
\begin{equation}
    \eta(\epsilon_\text{a}) = \frac{1}{\sigma(\epsilon_\text{a})}\int_0^{\epsilon_\text{a}}\sigma(\epsilon) d \epsilon,
\end{equation}
\begin{equation}
    \Bigg [\frac{d(\eta(\epsilon))}{d\epsilon} \Bigg]_{\epsilon=\epsilon_\text{D}} = 0.
\end{equation}
Upon crossing the onset strain of densification, the lattice becomes progressively `less efficient' at absorbing energy. Hence, a lattice structure that absorbs the most energy should exhibit a prolonged plastic deformation region, i.e., higher $\varepsilon_\text{D}$ and, as a result, higher SEA. We shall refer to the value of specific energy absorbed until $\varepsilon_\text{D}$ as SEA in this study, which will also be maximized.\\
The plateau stress, $\sigma_\text{pl}$  is defined as the average stress experienced by the fixed plate during the plastic deformation of the lattice, i.e., from the onset of yielding to the onset of densification, given by
\begin{equation}
    \sigma_\text{pl} = \frac{ \int_{\epsilon_\text{y}}^{\epsilon_\text{D}} \sigma(\epsilon) d \epsilon}{\int_{\epsilon_\text{y}}^{\epsilon_\text{D}} d \epsilon},
\end{equation}
where, the yield strain, $\epsilon_\text{y}$, is the strain corresponding to the value of $\sigma_\text{p}$. 
\subsection{Surrogate Model}\label{sec:surrogate_model}
Surrogate modeling has been a popular choice for solving optimization problems where generating the training dataset is cumbersome. These training datasets are usually compiled from experimental or simulation data, neither of which can be obtained on a large scale to build a large dataset. Surrogate modeling techniques, especially polynomial response surface methods, excel in this context due to their simplicity and robustness. The Thin Plate Spline Radial Basis Function (TPS \textendash RBF) method is adopted here to overcome the pitfalls associated with noise in higher-order polynomials between training points. The TPS-RBF is a surface-interpolation technique that approximates a 3D surface in our problem and computes intermediate values using radial basis function interpolation. The TPS-RBF response surface, $S(x,y)$ is defined as
\begin{equation}
    \hat{S}(x,y) = \sum_{i=1}^N w_i \phi(||(x,y) - (x_i,y_i)||) + a_0 + a_1x + a_2y,
\label{eq:example}
\end{equation}
where $N$ is the number of training points (or centers), $w_i$ are training weights, and $p(x,y)$ is a linear polynomial given by 
\begin{equation}
    p(x) = a_0 + a_1x + a_2y.
\end{equation}
$\phi$ is the kernel function, defined as $\phi(r)=r^2 \ln(r); \phi(0)=0$, where $r = ||(x,y)-(x_i,y_i)||$ is the euclidean distance between a query point $(x,y)$ and a training center $(x_i,y_i)$. Leave-One-Out Cross-Validation (LOOCV) was used to construct a surface robust to new, unseen data points. The quality of the TPS-RBF fit is estimated by computing the Mean Absolute Error (MAE) and the Root Mean Square Error (RMSE) given by
\begin{equation}
    \text{MAE} =\frac{1}{N} \sum_{i=1}^N e_i \quad \text{and} \quad \text{RMSE} =\sqrt{\frac{1}{N} \sum_{i=1}^N e_i^2}.
\end{equation}
The error, $e_i = |S_i(x_i,y_i) - \hat{S}_{-i}(x_i,y_i)|$, where, $S_i(x_i,y_i)$ is the actual target at the point $(x_i,y_i)$ and $\hat{S}_{-i}(x_i,y_i)$ is the prediction at $(x_i,y_i)$ from the TPS-RBF model trained without the $i-$th sample during cross-validation.
\subsection{Multi-objective Optimization: Goal Programming}
To find the best performing FG-BCC lattice according to our hypothesis, we need to maximize $\text{SEA}$ and $\text{Nu}$ and minimize $\sigma_\text{p}$ and $\Delta P$.  To achieve this target, we first construct surrogates for each of the four metrics individually. We then identify the required maximum or minimum of each of the four surrogates that can be obtained in the design space. Let the maximum values of SEA and $\text{Nu}$ be $\overline{\text{SEA}}$ and $\overline{\text{Nu}}$, respectively and the minimum values of $\sigma_\text{p}$ and $\Delta P$ be $\overline{\sigma_\text{p}}$ and $\overline{\Delta \text{P}}$, respectively.\\
The concept of goal programming requires that specific targets, or `goals,' are set for each objective. Here, the ideal goals are the previously found optimum values of our four metrics: $\overline{\text{SEA}}$, $\overline{\text{Nu}}$, $\overline{\sigma_\text{p}}$, and $\overline{\Delta \text{P}}$. This step generates a vector of ideal goals, G = [$\overline{\text{SEA}}$,$\overline{\text{Nu}}$,$\overline{\sigma_\text{p}}$,$\overline{\Delta \text{P}}$], which may not be practically attainable.
To find the optimal design ($d_1^*$, $d_2^*$) that is as close as possible to the ideal design, we allow deviations from the ideal-defined goals. For each objective, $j$, two types of deviations are possible:
\begin{itemize}
    \item Positive deviation ($d^+_j$): represents over-achievement of the goal
    \item Negative deviation ($d^-_j$): represents under-achievement of the goal
\end{itemize}
Each of these unwanted deviations is penalized for each unit of deviation that is observed. For example, in the case of maximizing Nu, we penalize a design, $j$, if $\text{Nu}_j < \overline{\text{Nu}} $, giving a positive penalty for $d^-_j$ and a zero penalty for $d^+_j$. In the case of minimizing $\Delta P$, we penalize a design, $j$, if $\Delta \text{P}_j > \overline{\Delta \text{P}}$, giving a positive penalty for $d^+_j$, and a zero penalty for $d^-_j$.
These penalties are user-defined and can be treated as hyperparameters. The essence of goal programming now reduces to minimizing the penalty-weighted sum of all deviations, Z, defined as:
\begin{equation}
    Z = \sum_{j=1}^4 \text{P}^+_j d^+_j + \text{P}^-_j d^-_j,
\end{equation}
where, j = {1,2,3,4} denotes the four objective functions of the problem and $\text{P}^+_j$ and $\text{P}^-_j$ are user-defined penalties for each objective function. Note that $\text{P}^+_j$, $\text{P}^-_j$, $d^+_j$ and $d^-_j$ are all non-negative quantities.
We use this method to find the optimal design ($d_1^*$, $d_2^*$) that achieves the best compromise with the lowest $Z$ value for a given set of penalties.
\section{Results}
Dynamic compression and forced convection simulations were performed for all 16 FG-BCC lattice designs, along with the ground BCC design, to obtain the values of SEA, $\sigma_\text{p}$, $\text{Nu}$, and $\Delta P$ for each lattice design. The combination of all the design points, $x_i = (d_1, d_2)_i$ with its corresponding $y_i$, where $i$ represents the $i^\text{th}$ lattice design, forms the training data, $(x_i, y_i)$ ($y_i$ being the metric of interest) for running the surrogate algorithm to construct surrogates representing the individual polynomial surfaces in the design space. \Cref{tab:ground_simulation_results} summarises the simulation results for the ground structure, which will serve as the basis for comparison.
\begin{table*}[t]
\centering
\caption{Simulation results of the ground BCC lattice structure}
\label{tab:ground_simulation_results}
\begin{tabular*}{\textwidth}{@{\extracolsep{\fill}}cccccccc}
\toprule
Lattice ID & SEA (kJ/kg) & $\sigma_\text{p}$ (MPa) & $\sigma_\text{pl}$ (MPa) & Nu & $\Delta P$ (Pa) & $C_\text{f}$ & $\beta$ \\
\midrule
0 & 8.68 & 6.19 & 10.16 & 5587.30 & 632.04 & 4.59 & 3362.92 \\
\bottomrule
\end{tabular*}
\end{table*}
The simulation results for the remaining 16 design points are shown in \Cref{app:simulation_values}.

\subsection{Surrogate Modeling}\label{sec:surrogate_model_results}
\begin{figure*}[t]
    \centering

    \begin{subfigure}{0.48\textwidth}
        \centering
        \includegraphics[width=0.95\linewidth]{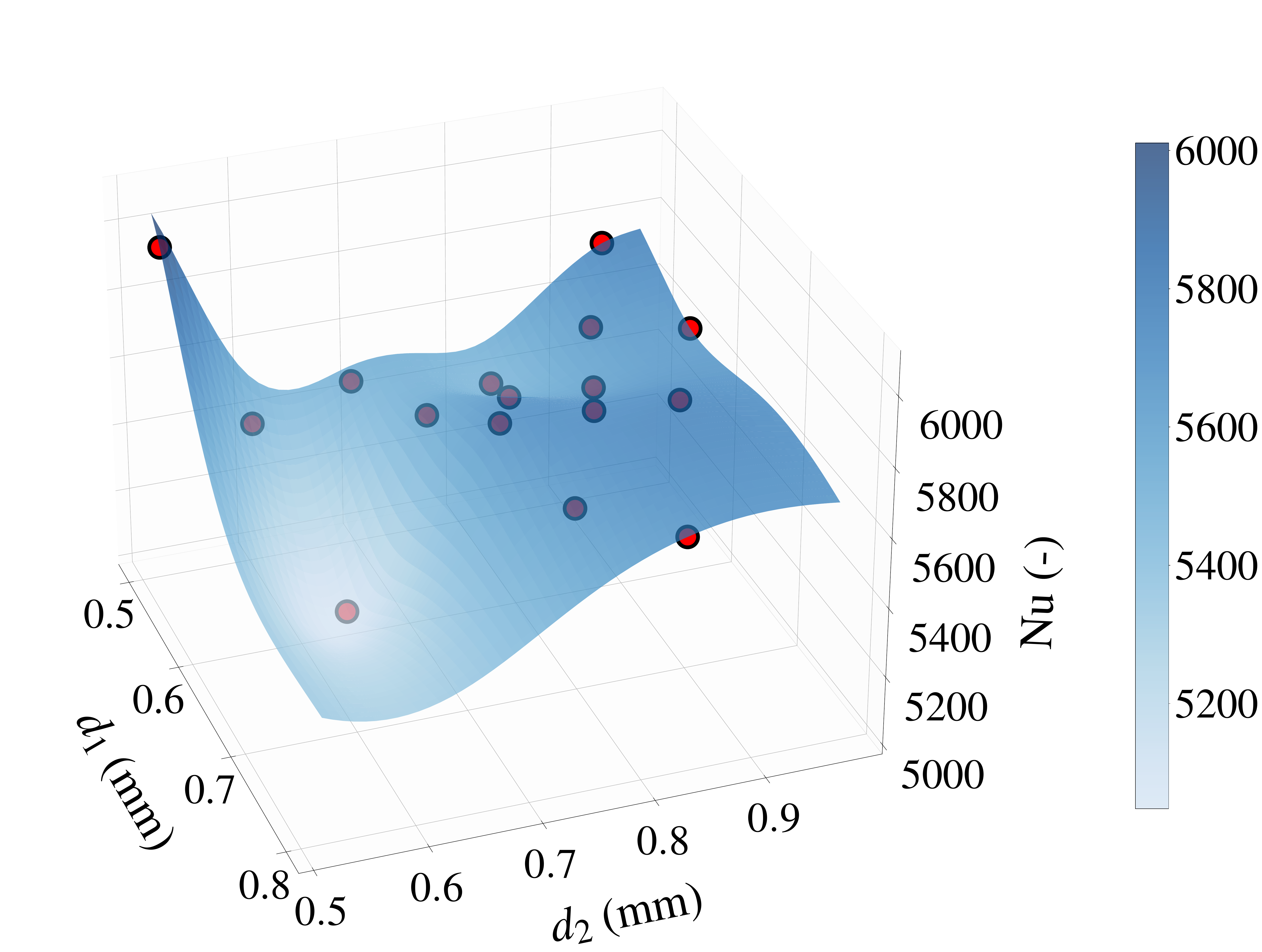}
        \caption{}
        \label{fig:Nu_surrogate}
    \end{subfigure}
    \hfill
    \begin{subfigure}{0.48\textwidth}
        \centering
        \includegraphics[width=\linewidth]{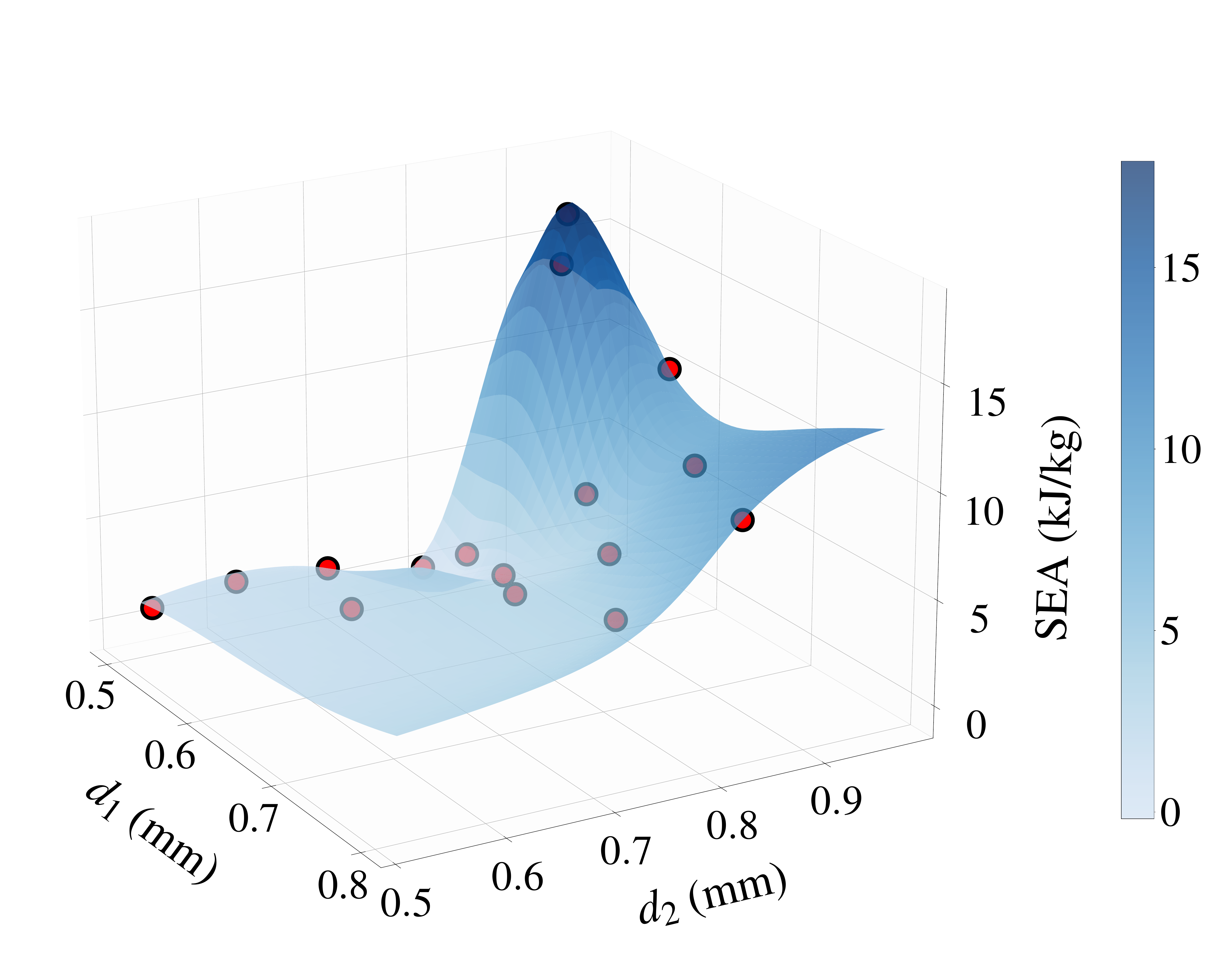}
        \caption{}
        \label{fig:SEA_surrogate}
    \end{subfigure}

    \begin{subfigure}{0.48\textwidth}
        \centering
        \includegraphics[width=0.95\linewidth]{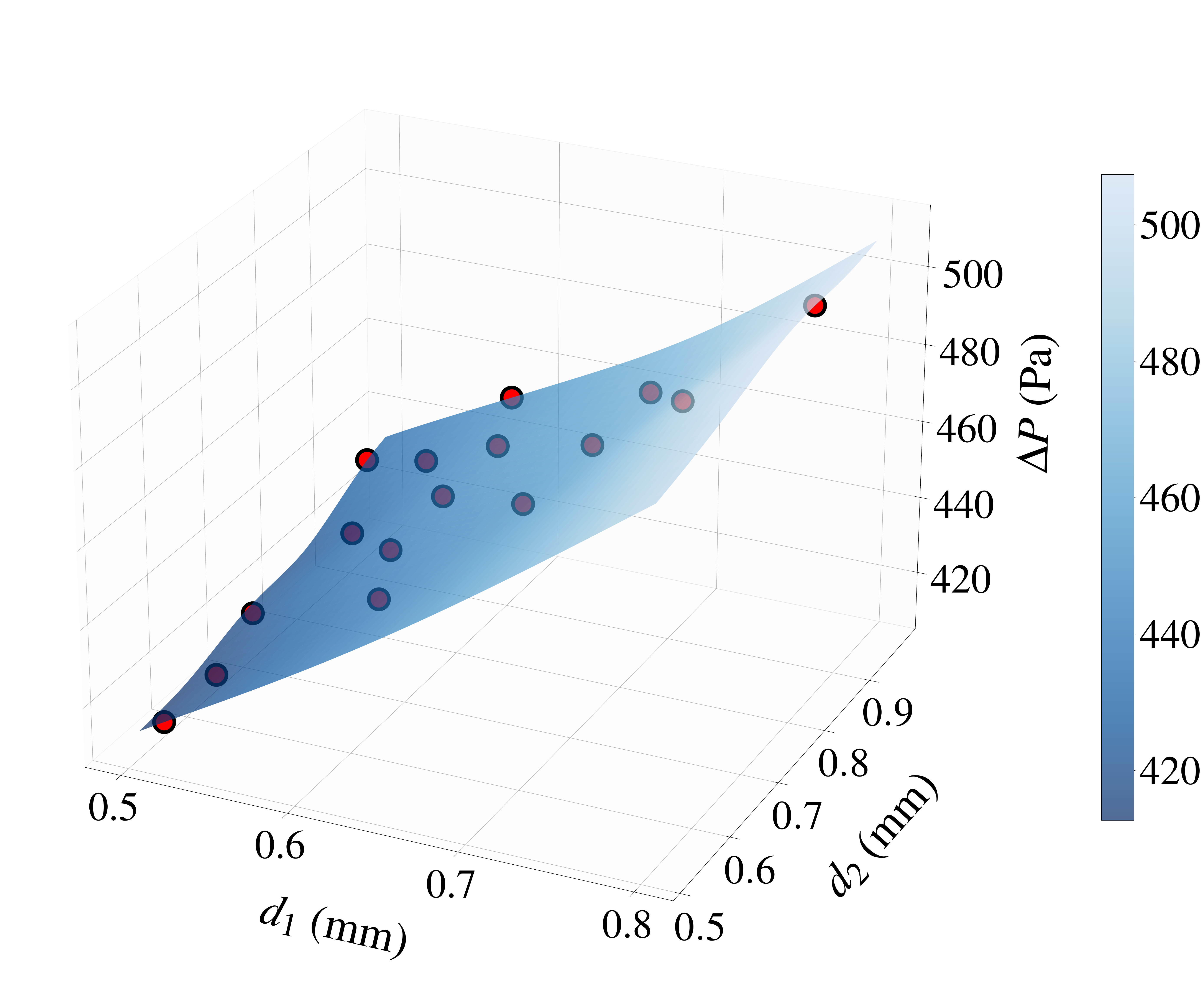}
        \caption{}
        \label{fig:DelP_Surrogate}
    \end{subfigure}
    \hfill
    \begin{subfigure}{0.48\textwidth}
        \centering
        \includegraphics[width=\linewidth]{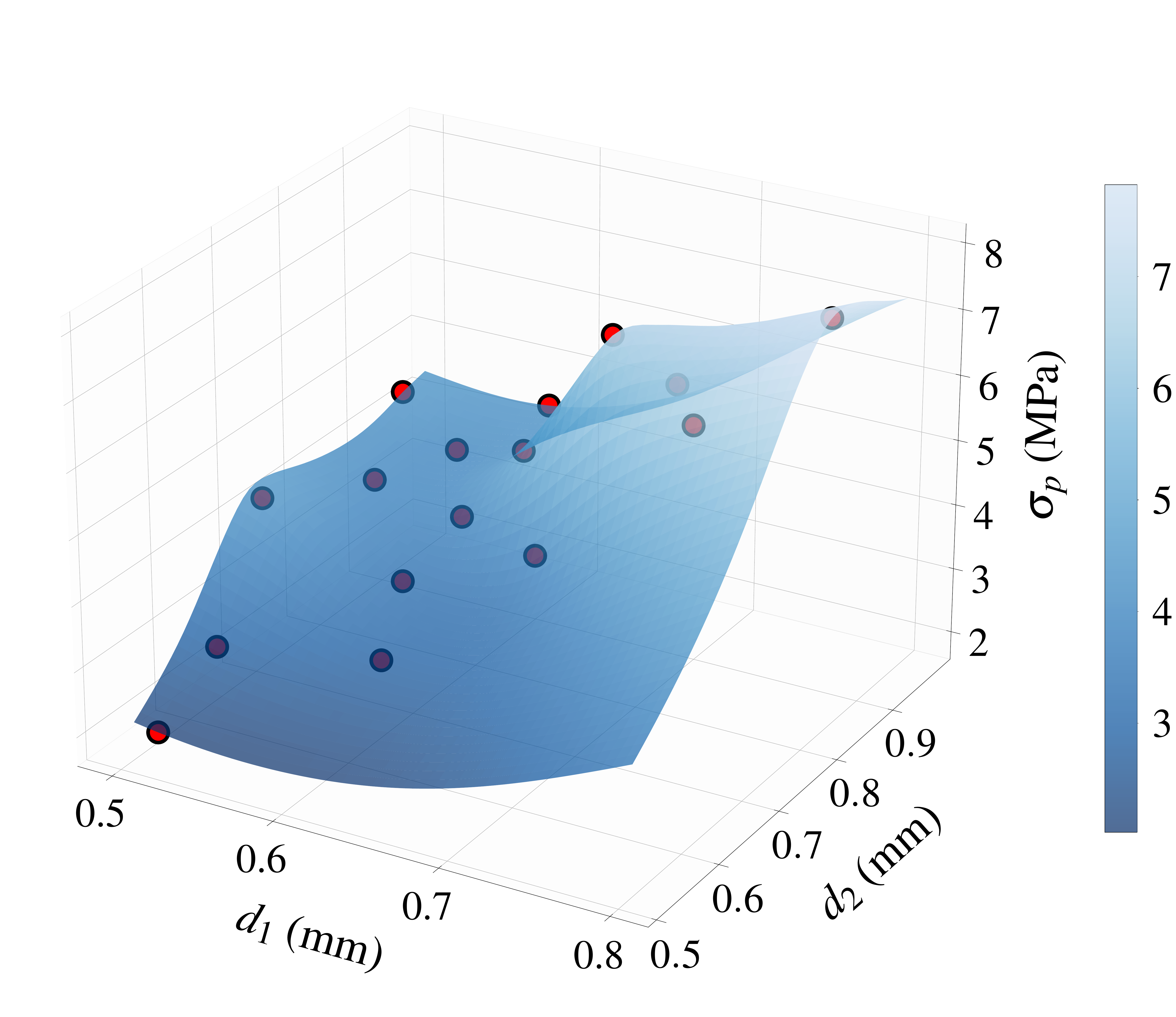}
        \caption{}
        \label{fig:sigma_p_Surrogate}
    \end{subfigure}

    \caption{Compilation of surface plots of the (a) Nusselt number, (b) maximum specific energy absorption, (c) pressure drop, and (d) peak stress surrogates, constructed using the 16 training points (red) within the design space}
    \label{fig:2x2_surrogate_setup}
\end{figure*}

TPS-RBF surrogates were constructed using the methodology described in \Cref{sec:surrogate_model} for each metric of interest: Nusselt number (Nu), Specific Absorption Energy (SEA), pressure drop ($\Delta P$), and peak stress ($\sigma_\text{p}$). The corresponding surfaces are shown in \Cref{fig:2x2_surrogate_setup}.

The R$^2$ value of all these surrogates is 1.0, and the MAE and RMSE of each of these surrogates are shown in \Cref{tab:surrogate_errors}.

\begin{table}[htbp]
    \centering
    \normalsize
    \caption{Surrogate model errors}
    \label{tab:surrogate_errors}
    \begin{tabular}{ccc}
        \toprule
        Surrogate & MAE & RMSE \\
        \midrule
        Nusselt number & 3.98e-13 & 6.82e-13 \\
        SEA & 2.04e-14 & 2.90e-14 \\
        Pressure Drop ($\Delta P$) & 5.33e-14 & 6.19e-14 \\
        Peak Stress ($\sigma_\text{p}$) & 2.03e-15 & 2.85e-15 \\
        \bottomrule
    \end{tabular}
\end{table}
These surrogates indicate that the relationship between the metrics and the design variables is highly nonlinear, except for pressure drop, which shows an almost linear trend. The linear nature of the pressure drop function with respect to the diameter values can be explained by the Darcy-Weisbach equation for a flow in a cylindrical pipe, given by
\begin{equation}
    \Delta {P} = f \times (L/d_\text{H}) \times \frac{\rho \overline{u}^2}{2},
\end{equation}
where $f$ is the friction factor, $L$ is the characteristic flow length, $d_\text{H}$ is the hydraulic diameter, and $\overline{u}$ is the mean flow velocity. In our case, for a fixed value of $d_2$, as $d_1$ increases, the resultant pore space through which the fluid flows, decreases. As a result, a reduction in the hydraulic diameter, $d_\text{H}$, is seen. Hence, as $d_1$ increases, $d_\text{H}$ decreases, resulting in a linear increase in the pressure drop, $\Delta P$.

Regarding the Nusselt number surrogate, it can be observed that lattice designs with the smallest $d_1$ and $d_2$ values (or the highest $d_3$ value, due to the relative density constraint) exhibit the highest Nusselt number values. We also observe that the Nusselt number is more sensitive to $d_2$ than to $d_1$. Specific Energy Absorption (SEA), on the other hand, is higher for designs with a combination of a lower $d_1$ and a higher $d_2$, and the increase in SEA is rapid near the corner values. The peak stress surrogate shows a minimum for designs with the smallest $d_1$ and $d_2$ values. 
These observations precisely match the hypothesis, as described in \Cref{section:introduction}. Lattice designs with thinner struts (lower $d_1$ and $d_2$ values) at the top exhibit the lowest peak stresses (under impact) and pressure drop (under forced convection), as expected. Additionally, these lattices, with a significantly larger $d_3$ value (thicker struts at the bottom), exhibit the highest Nusselt number, as expected, due to the increased contact surface area.

\subsection{Goal Programming}
Now that we have the surrogates ready, the maximum values of Nu and SEA surrogates, along with the minimum values of $\Delta P$ and $\sigma_\text{p}$, are found using the L-BFGS method. These values will serve as targets, or `goals,' for the multi-objective optimization problem. The identified goals for our problem are given in \Cref{tab:multiObjOpt_Goals}.

\begin{table}[htbp]
    \centering
    \normalsize
    \caption{Goals for multi-objective optimization}
    \label{tab:multiObjOpt_Goals}
    \begin{tabular}{c c c c}
    \hline
    \rule{0pt}{2.5ex} 
    $\overline{\text{Nu}}$ & $\overline{\text{SEA}}$ (kJ/kg) & $\overline{\Delta P}$ (Pa) & $\overline{\sigma_\text{p}}$ (MPa)\\
    \hline
    6052.59 & 18.06 & 411.83 & 1.99\\
    \hline
    \end{tabular}
\end{table}

\begin{figure}[htbp]
    \centering
    \includegraphics[width=0.6\linewidth]{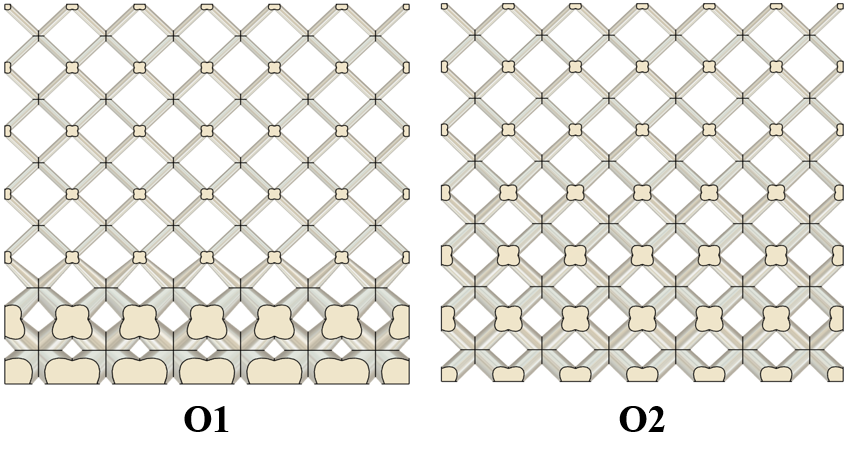}
    \caption{The cross-section of O1 and O2 lattice geometries}
    \label{fig:O1_O2_geometry}
\end{figure}

The goal programming algorithm was run with a penalty of 1 (or equal weightage to all metrics) for all goal deviations, and the results are shown in \Cref{tab:GP_result_1}.
\begin{table}[htbp]
\centering
\caption{Goal programming result with equal penalties}
\label{tab:GP_result_1}
\begin{tabular}{c c c c c c}
\hline
$d_1^*$ (mm) & $d_2^*$ (mm) & Nu & SEA (kJ/kg) & $\Delta P$ (Pa) & $\sigma_\text{p}$ (MPa) \\
\hline
0.503 & 0.520 & 6052.59 & 1.16 & 411.83 & 1.99 \\
\hline
\end{tabular}
\end{table}
Here, $d_1^*$ and $d_2^*$ are the optimal diameter values. While this optimal design (called O1 hereafter) achieves the best thermal performance and has very low $\sigma_\text{p}$ and $\Delta P$ values, its SEA value is considerably lower than the target and even the ground structure. This can be attributed to the dominance of the lattice designs, which have the lowest $d_1$ and $d_2$ values across three of the four metrics. To achieve a more meaningful tradeoff, we increase the penalty for SEA deviation to 250 and rerun the goal programming algorithm; the results are shown in \Cref{tab:GP_result_2}.
\begin{table*}[htbp]
\centering
\caption{Goal programming result with a higher penalty for SEA deviation}
\label{tab:GP_result_2}
\begin{tabular}{c c c c c c}
\hline
$d_1^*$ (mm) & $d_2^*$ (mm) & Nu & SEA (kJ/kg) & $\Delta P$ (Pa) & $\sigma_\text{p}$ (MPa) \\
\hline
0.539 & 0.919 & 5679.71 & 17.22 & 437.31 & 4.11 \\
\hline
\end{tabular}
\end{table*}
This design (called O2 hereafter) shows a fairly balanced tradeoff in Nu for gain in SEA. Although the value of Nu is significantly lower than the set goal, it is comparable to the ground structure, while all the other metrics have shown substantial improvements. A much more comprehensive comparison of the designs O1 and O2 with the ground structure is given in \Cref{tab:ground_to_optima_comparison}. The generated O1 and O2 lattice geometries are shown in \Cref{fig:O1_O2_geometry}.
\begin{table*}[htbp]
\centering
\caption{Comparing the performance of optimal lattice designs with the ground lattice}
\label{tab:ground_to_optima_comparison}
\begin{tabular}{c c c c c c c c}
\hline
Lattice ID & $d_1^*$ (mm) & $d_2^*$ (mm) & Nu & SEA (kJ/kg) & $\Delta P$ (Pa) & $\sigma_\text{p}$ (MPa) & $\beta$ \\
\hline
0 (Ground Structure) & 0.780 & 0.780 & 5587.30 & 8.68 & 632.04 & 6.19 & 3362.92 \\ \hline
O1 & 0.500 & 0.520 & 6052.59 & 1.16 & 411.83 & 1.99 & 4202.08 \\
\% Improvement & - & - & 8.33 & -88.18 & 34.84 & 66.65 & 24.95 \\ \hline
O2 & 0.539 & 0.919 & 5679.71 & 17.22 & 437.31 & 4.11 & 3865.09 \\
\% Improvement & - & - & 1.65 & 114.77 & 30.81 & 35.95 & 14.93 \\
\hline
\end{tabular}
\end{table*}
Clearly, the design O2 is better than O1; despite a minimal improvement ($\approx$ 2 \%) in Nu, it still exhibits an overall thermal performance improvement ($\approx$ 15 \%), as indicated by $\beta$, due to the lower pressure drop.
\subsection{O1 vs. O2 Lattice}
To evaluate the reliability of our framework, impact and forced-convection simulations were performed on both lattice designs, O1 and O2. The simulation results are presented in \Cref{tab:GP_Simulation_comparison}, indicating that the simulation results align well with the goal programming results.
\begin{table*}[htbp]
\centering
\caption{Validation of the goal programming results with simulation results}
\label{tab:GP_Simulation_comparison}
\begin{tabular}{c c c c c c c}
\hline
Lattice ID & $d_1^*$ (mm) & $d_2^*$ (mm) & Nu & SEA (kJ/kg) & $\Delta P$ (Pa) & $\sigma_\text{p}$ (MPa) \\
\hline
O1 -- GP & 0.500 & 0.520 & 6052.59 & 1.16 & 411.83 & 1.99 \\
O1 -- Simulation & 0.500 & 0.520 & 6075.59 & 1.214 & 417.31 & 2.352 \\
\% Error & - & - & 0.38 & 4.66 & 1.33 & 18.19 \\ \hline
O2 -- GP & 0.539 & 0.919 & 5679.71 & 17.22 & 437.31 & 4.11 \\
O2 -- Simulation & 0.539 & 0.919 & 5664.37 & 16.447 & 443.39 & 4.203 \\
\% Error & - & - & 0.27 & 4.70 & 1.39 & 2.26 \\
\hline
\end{tabular}
\end{table*}

\subsection{PIGOs Correlation Analysis}
This subsection presents the Pearson correlation analysis of the raw diameter values and various PIGOs proposed as surrogate design variables. Three scalar operators are derived from the four diameter control values $(d_0, d_1, d_2, d_3)$ and the area porosity function $\varphi(D)$ from \Cref{eq:area_porosity_calculation}, while six more operators are computed directly from the triangulated surface mesh (STL) of each FG-BCC lattice design, requiring no analytical assumptions 
about strut geometry. Each operator encodes a specific geometric feature that 
is hypothesized to govern one or more performance metrics. A summary of all nine operators, their definitions, and their physical interpretations is given below:

\begin{itemize}         
    \item \textbf{Diameter gradient curvature, $\nabla^2 D$} is the second finite difference of 
            the diameter profile:
            \begin{equation}
                \nabla^2 D = \frac{d_3 - 2d_2 + d_1}{\Delta z^2}, \quad \Delta z = H/3,
                \label{eq:grad_curvature}
            \end{equation}
            where $H$ is the total height of the lattice. $\nabla^2 D > 0$ indicates a convex gradient, where diameter increases slowly through the upper zones then sharply near the heat source, producing compound-lattice behavior and poor SEA, as observed in FG-BCC-3. $\nabla^2 D \approx 0$ corresponds to a linear gradient and the progressive layer-by-layer collapse characteristic of the O2 lattice. 

    \item \textbf{Porosity gradient, $\nabla\varphi$} is the normalized difference in area 
            porosity between the top and bottom layers:
            \begin{equation}
                \nabla\varphi = \frac{\varphi(d_0) - \varphi(d_3)}{H}
                \label{eq:porosity_gradient}
            \end{equation}
           A steep positive $\nabla\varphi$ (open top, dense bottom) simultaneously reduces flow resistance and increases conductive surface area, maximizing both $\text{Nu}$ and minimizing $\sigma_p$ in the training data.

    \item \textbf{Impingement surface porosity, $\Omega$} is the ratio of the top-layer porosity 
            to the mean porosity across all layers:
            \begin{equation}
                \Omega = \frac{\varphi(d_0)}{\bar{\varphi}}
                \label{eq:impingement_effectiveness}
            \end{equation}
            $\Omega > 1$ indicates a more open top layer than average, allowing the incoming jet to 
            penetrate deep into the lattice and distribute cooling uniformly. $\Omega < 1$ causes flow 
            stagnation and bypass, as seen to be responsible for the O1 lattice's elevated chip temperature despite its high $\text{Nu}$. As shown in \Cref{app:pigo_tables}, $\Omega$ achieves the strongest single correlation in the entire dataset: $r = -0.965$ with $\Delta P$.

    \item \textbf{Solid-fluid interface area, $A_\mathrm{sf}$} is the total surface area of the triangulated mesh,
    \begin{equation}
        A_\mathrm{sf} = \sum_f a_f, 
    \end{equation}
    where $a_f$ is the area of each triangular face. Unlike the analytical area porosity expression, $A_\mathrm{sf}$ captures the actual strut-joint geometry, including overlap regions at nodes. It is the primary geometric driver of convective heat transfer, as a larger solid-fluid interface directly increases the area available for thermal exchange.

    \item \textbf{Normalised mean-curvature integral, $\kappa_\mathrm{int}$} is the surface-averaged absolute mean curvature,
    \begin{equation}
        \kappa_\mathrm{int} = \frac{1}{A_\mathrm{sf}} \int_S |C|\,\mathrm{d}A
    \end{equation}
    where $C$ is the discrete mean curvature estimated at each mesh vertex. High $\kappa_\mathrm{int}$ indicates sharp strut-to-node transitions across the mesh surface, which are geometric stress-concentration sites governing the initial peak stress $\sigma_p$. This operator is directly inspired by the curvature-integral operators of \cite{khan2025physics}.

    \item \textbf{Principal moment of inertia ratio, $\eta_I$} is the ratio of the smallest to largest eigenvalue of the solid inertia tensor, $\eta_I = I_\mathrm{min} / I_\mathrm{max}$, computed assuming uniform density throughout the mesh. For a perfectly symmetric uniform lattice, $\eta_I = 1$. Functionally graded lattices with heavier bottom zones have $\eta_I < 1$, encoding axial mass asymmetry in a way that is independent of any parameterization of the diameter profile.

    \item \textbf{Convex-hull packing ratio, $\chi$} is the ratio of the solid mesh volume to the volume of its convex hull, $\chi = V_\mathrm{solid} / V_\mathrm{hull}$. This is the three-dimensional analog of the solidity descriptor used in two-dimensional shape analysis \cite{khan2025physics}. Higher $\chi$ indicates a denser, less porous structure; lower $\chi$ indicates a more open lattice. $\chi$ encodes how efficiently the solid fills its bounding envelope, which shifts monotonically with gradation steepness.

    \item \textbf{Normalized axial second moment, $J_z$} is the second moment of the solid volume about the center of mass along the gradation axis, normalized by the total volume and height squared,
    \begin{equation}
        J_z = \frac{1}{V H^2} \int_V (z - z_\mathrm{cm})^2\,\mathrm{d}V
    \end{equation}
    computed by voxelising the mesh. $J_z$ distinguishes end-loaded from center-loaded mass distributions: a high $J_z$ indicates mass concentrated near the top and bottom ends, while a low $J_z$ indicates mass concentrated near the center. This information is complementary to $\eta_I$, which mixes all three principal axes.

    \item \textbf{Thermally active zone surface fraction, $\tau$} is the fraction of the total mesh surface area located within the second, third, and fourth layers from the impingement face of the lattice,
    \begin{equation}
        \tau = \frac{A_\mathrm{mid}}{A_\mathrm{sf}}
    \end{equation}
    where $A_\mathrm{mid}$ is the surface area of all triangular faces whose centroid falls within $[H/6,\, 4H/6]$ from the impact end. In the jet impingement configuration used here, the incoming jet core occupies the upper third, deflects laterally through the middle third, and stagnates near the bottom. $\tau$ therefore encodes how much solid-fluid interface is located where the deflected jet is still cold in the zone where convective exchange is most productive.
\end{itemize}

\noindent Operator values for all 17 lattice designs are provided in \Cref{tab:pigo_values}. \Cref{tab:pearson_correlations} reports the Pearson correlation coefficients between design variables, analytical operators, and STL-derived operators against simulation outputs across the 17 FG-BCC lattice designs.

\begin{table*}[htbp]
\centering
\caption{Pearson correlation coefficients across 17 FG-BCC lattice designs. Bold values indicate $|r| \geq 0.70$.}
\label{tab:pearson_correlations}
\begin{tabular}{lcccc}
\toprule
Variable & SEA & $\sigma_p$ & Nu & $\Delta P$ \\
\midrule
$d_1$ & 0.15 & \textbf{0.817} & 0.065 & \textbf{0.773} \\
$d_2$ & \textbf{0.721} & 0.510 & 0.193 & 0.199 \\
\midrule
$\nabla^2 D$ & $-$0.684 & $-$\textbf{0.713} & $-$0.157 & $-$0.493 \\
$\nabla\varphi$ & $-$0.506 & $-$\textbf{0.826} & $-$0.151 & $-$\textbf{0.868} \\
$\Omega$ & $-$0.312 & $-$0.585 & $-$0.081 & $-$\textbf{0.965} \\
\midrule
$A_\mathrm{sf}$ & 0.458 & \textbf{0.761} & $-$0.036 & 0.616 \\
$\kappa_\mathrm{int}$ & $-$0.067 & $-$0.217 & $-$0.262 & $-$0.590 \\
$\eta_I$ & $-$0.188 & 0.365 & $-$0.310 & 0.680 \\
$\chi$ & $-$0.485 & $-$\textbf{0.798} & $-$0.048 & $-$0.658 \\
$J_z$  & 0.417 & \textbf{0.809} & 0.032 & \textbf{0.794} \\
$\tau$ & 0.329 & 0.671 & $-$0.261 & 0.573 \\
\bottomrule
\end{tabular}
\end{table*}

\section{Discussion}
\subsection{Impact Performance} \label{subsec:impact_discussion}
To better understand the reason behind the improvement in the impact performance of the O2 lattice over the ground lattice, the stress-strain plots for both structures are studied, as shown in \Cref{fig:ground_stressStrain} and \Cref{fig:O2_stressStrain}, which reveal that: 
\begin{figure}[htbp]
    \centering
    \begin{subfigure}{0.48\textwidth}
        \centering
        \includegraphics[width=\linewidth]{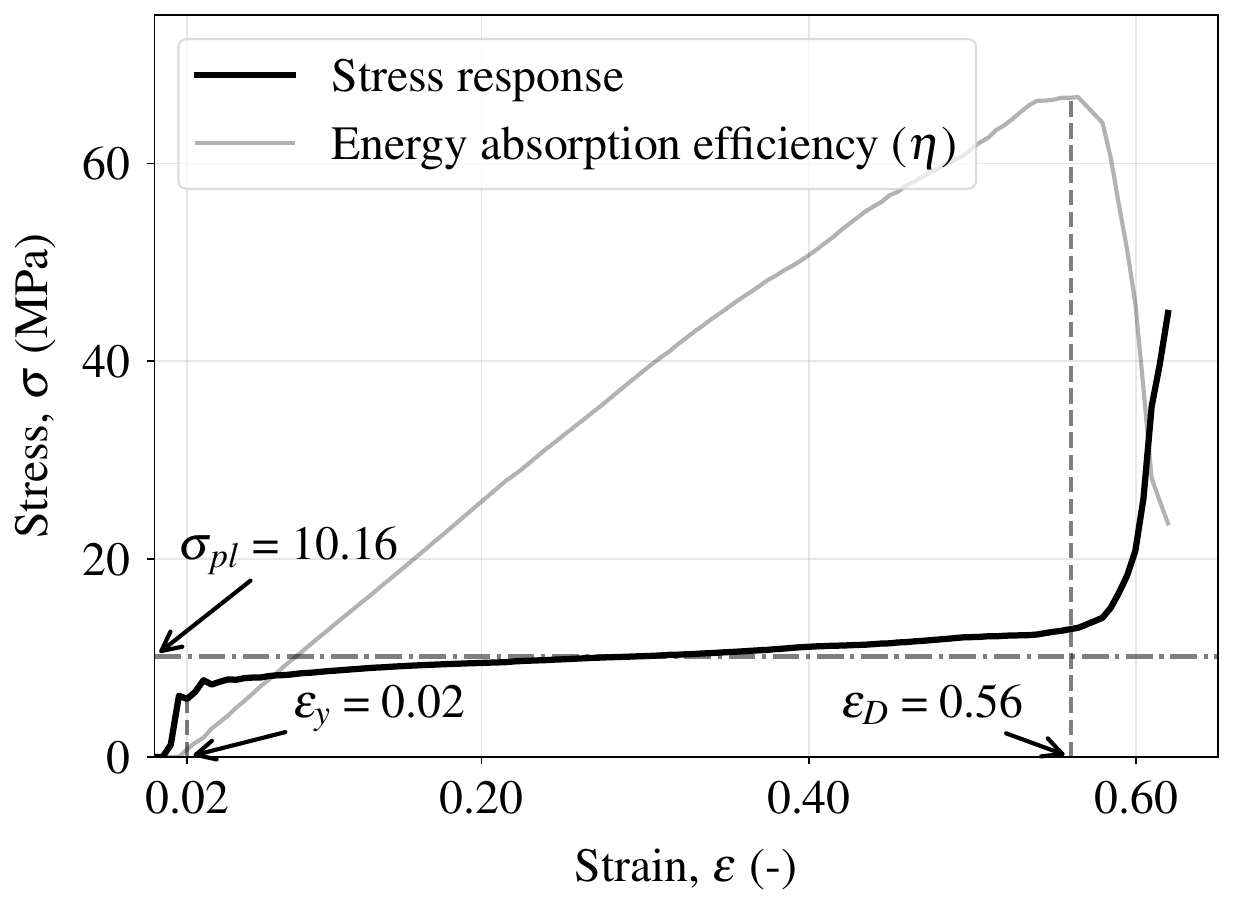}
        \caption{Ground lattice} 
        \label{fig:ground_stressStrain}
    \end{subfigure}
    \hfill
    \begin{subfigure}{0.48\textwidth}
        \centering
        \includegraphics[width=\linewidth]{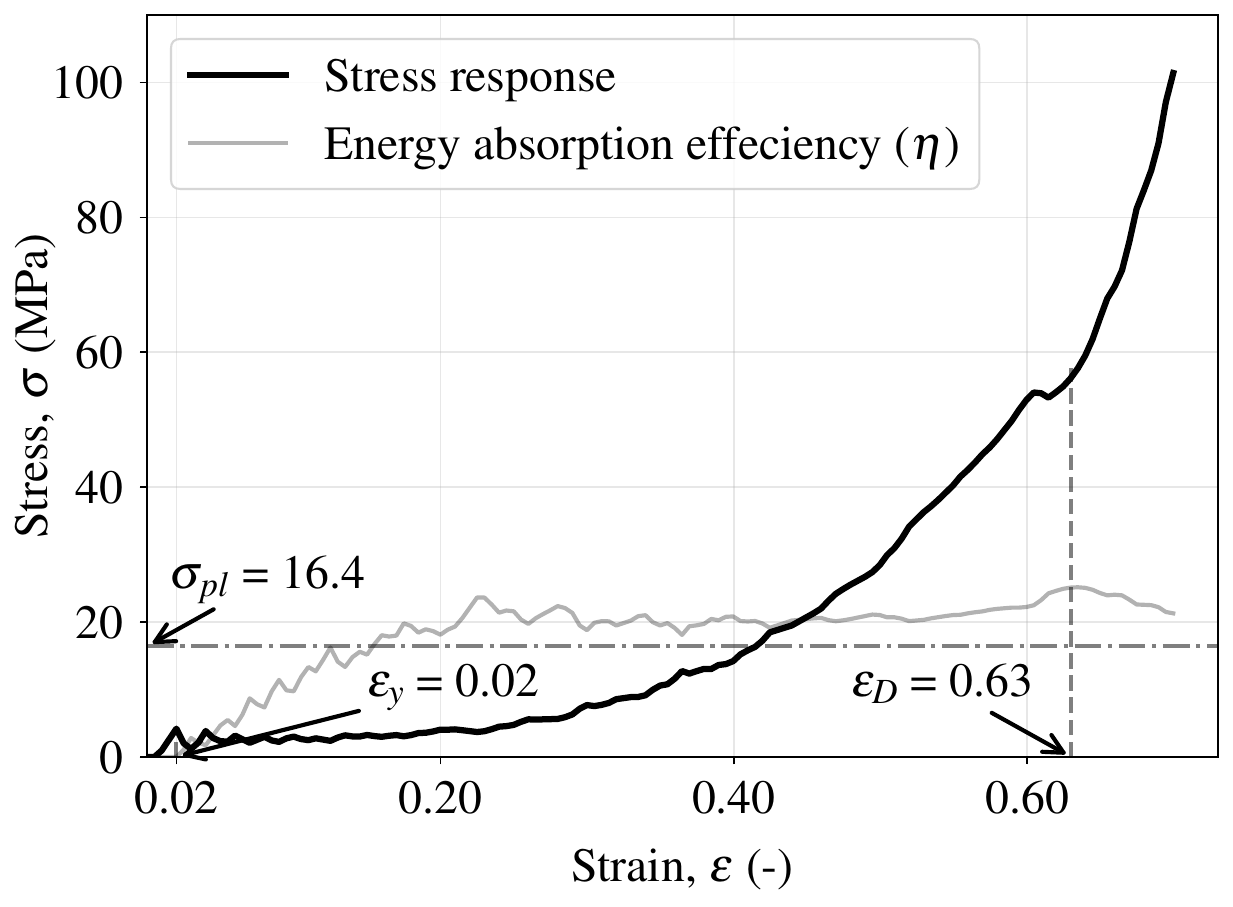}
        \caption{Optimized O2 lattice} 
        \label{fig:O2_stressStrain}
    \end{subfigure}

    \caption{Comparison of stress–strain response and energy absorption efficiency. The O2 lattice shows delayed densification, a steadier increase in stress, higher energy absorption, and a lower initial peak stress compared to the ground lattice.}
    \label{fig:stress_strain_comparison}
\end{figure}

\begin{figure*}[t]
    \centering
    \begin{subfigure}{\linewidth}
        \centering
        \includegraphics[width=\linewidth]{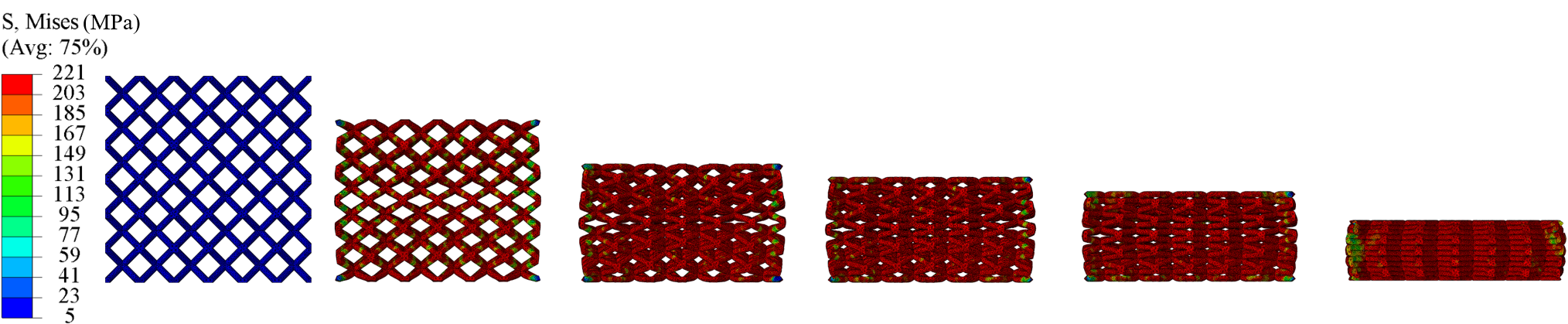}
        \put(-407,0){$\varepsilon=0$}
        \put(-348,0){$\varepsilon=0.165$}
        \put(-275,0){$\varepsilon=0.360$}
        \put(-206,0){$\varepsilon=0.435$}
        \put(-130,0){$\varepsilon=0.525$}
        \put(-52,0){$\varepsilon=0.675$}
        \caption{}
        \label{fig:ground_deformation_pattern}
    \end{subfigure}

    \begin{subfigure}{\linewidth}
        \centering
        \includegraphics[width=\linewidth]{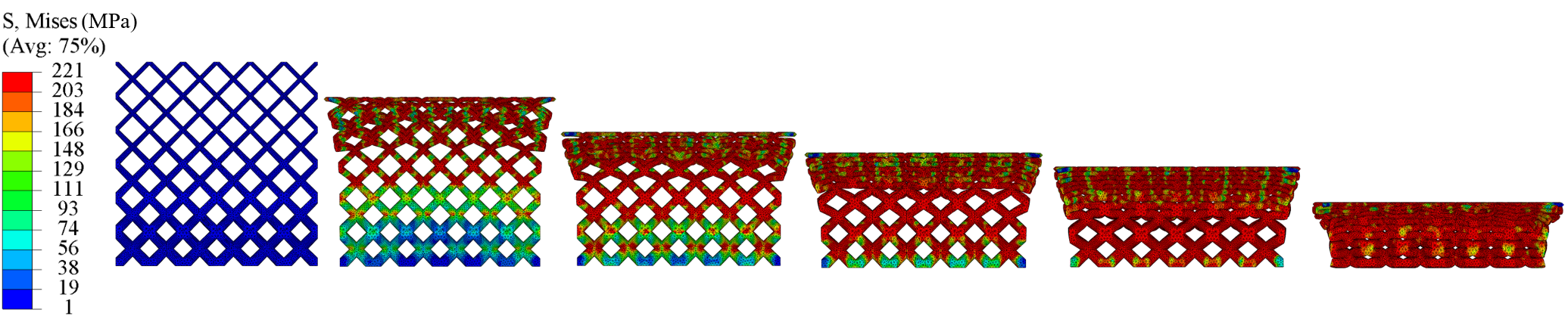}
        \put(-407,0){$\varepsilon=0$}
        \put(-348,0){$\varepsilon=0.165$}
        \put(-279,0){$\varepsilon=0.360$}
        \put(-205,0){$\varepsilon=0.435$}
        \put(-135,0){$\varepsilon=0.525$}
        \put(-57,0){$\varepsilon=0.675$}
        \caption{}
        \label{fig:O2_deformation_pattern}
    \end{subfigure}

    \caption{Comparative deformation patterns between (a) ground BCC showing uniform stress distribution and consistent compression, and (b) O2 lattice showing stress localization and sequential densification at various strain levels}
    \label{fig:combined_lattice_study}
\end{figure*}

\begin{itemize}
    \item The stress during plastic deformation is almost constant and smooth in the case of the ground structure, but it is continuously increasing with fluctuations in the case of the O2 structure. This can be attributed to the increasing strut diameter along the impact direction in the O2 structure, as opposed to the uniform-diameter struts in the ground structure.
    \item $\eta$ is almost linearly increasing in the case of ground structure, but it is highly fluctuating, and increases slowly till $\varepsilon_\text{D}$ in the case of the O2 structure.
    \item The onset strain of densification ($\varepsilon_\text{D}$) is higher in the O2 structure than the ground structure, showing that the O2 structure loses its "usefulness" much later than the ground structure, as $\varepsilon_\text{D}$ is defined as the strain at the global maximum of the energy absorption efficiency curve.
    \item The O2 structure has a lower initial peak stress as compared to the ground structure.
\end{itemize}

The ground structure exhibits a uniform and steady collapse of all layers together, as shown in \Cref{fig:ground_deformation_pattern}. The O2 structure, on the other hand, as shown in \Cref{fig:O2_deformation_pattern}, has progressive collapse. The collapse of a layer begins only when the collapse of the previous layer has been completed. This layer-by-layer collapse causes the stress values to increase at regular intervals, as fresh struts that have undergone negligible plastic deformation at lower layers are always available, especially in the initial stages of deformation.
\subsection{Thermal Performance}
\label{subsec:thermal_discussion}
We have so far seen that the O1 lattice has the highest Nusselt number, lowest pressure drop, and also the highest effectiveness, $\beta$. Increasing the strut diameter decreases porosity, creating two competing effects. First, the flow resistance increases, reducing the mass flow rate of the cold air and thus decreasing overall convection. Second, the total surface area in contact with the cold air increases, providing more area for heat transfer and thereby increasing convection. The net thermal performance, therefore, depends on which of these effects is dominant.

\begin{figure*}[t]
    \centering
    \includegraphics[width=\linewidth]{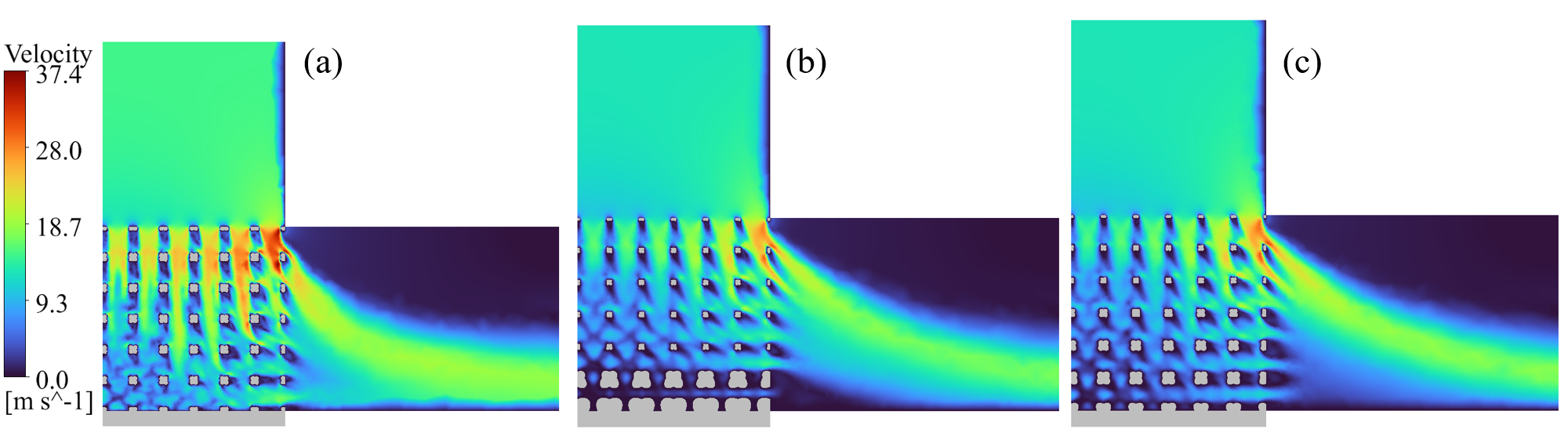}
    \caption{Velocity contours are shown on the mid-plane for three types of lattices: (a) the ground lattice, (b) the O1 lattice, and (c) the O2 lattice. The flow patterns in the ground BCC lattice are similar to those in the O2 lattice. In contrast, the O1 lattice exhibits minimal flow in its bottommost layer, where tight pore channels are present}
    \label{fig:fluent_velocity contours}
\end{figure*}

\begin{figure*}[t]
    \centering
    \includegraphics[width=\linewidth]{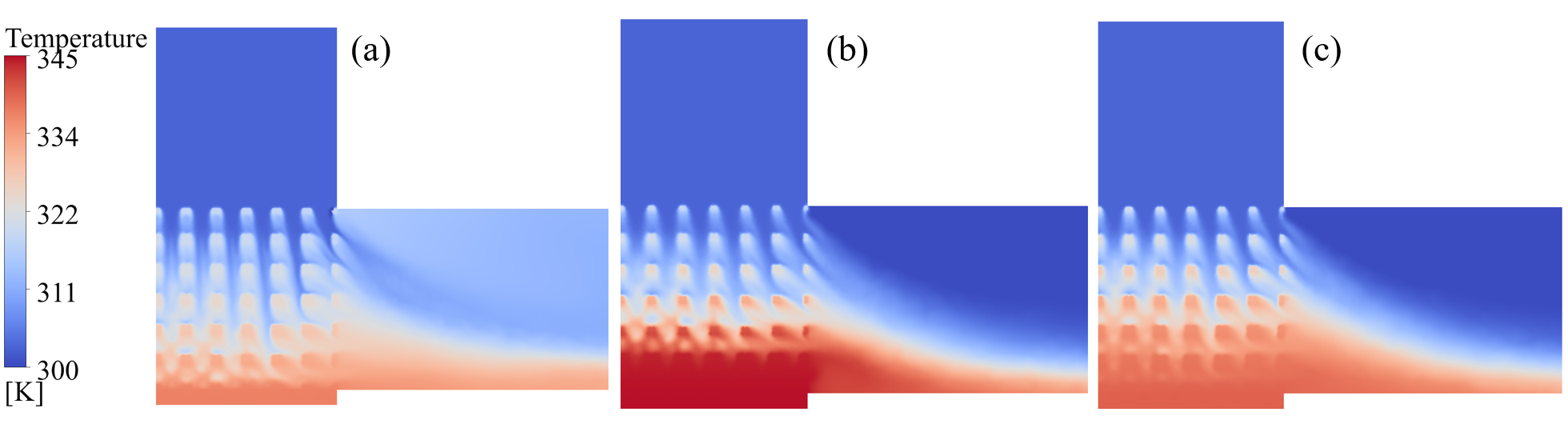}
    \caption{The temperature distributions on the mid-plane are illustrated for three scenarios: (a) the ground lattice, (b) the O1 lattice, and (c) the O2 lattice. The chip temperature is notably higher in the O1 lattice case. In contrast, both the ground lattice and the O2 lattice yield significantly lower chip temperatures, with the ground lattice exhibiting the lowest temperature of all, albeit in close proximity to the O2 lattice}
    \label{fig:fluent_temperature_contours}
\end{figure*}

\begin{figure*}[t]
    \centering
    \includegraphics[width=\linewidth]{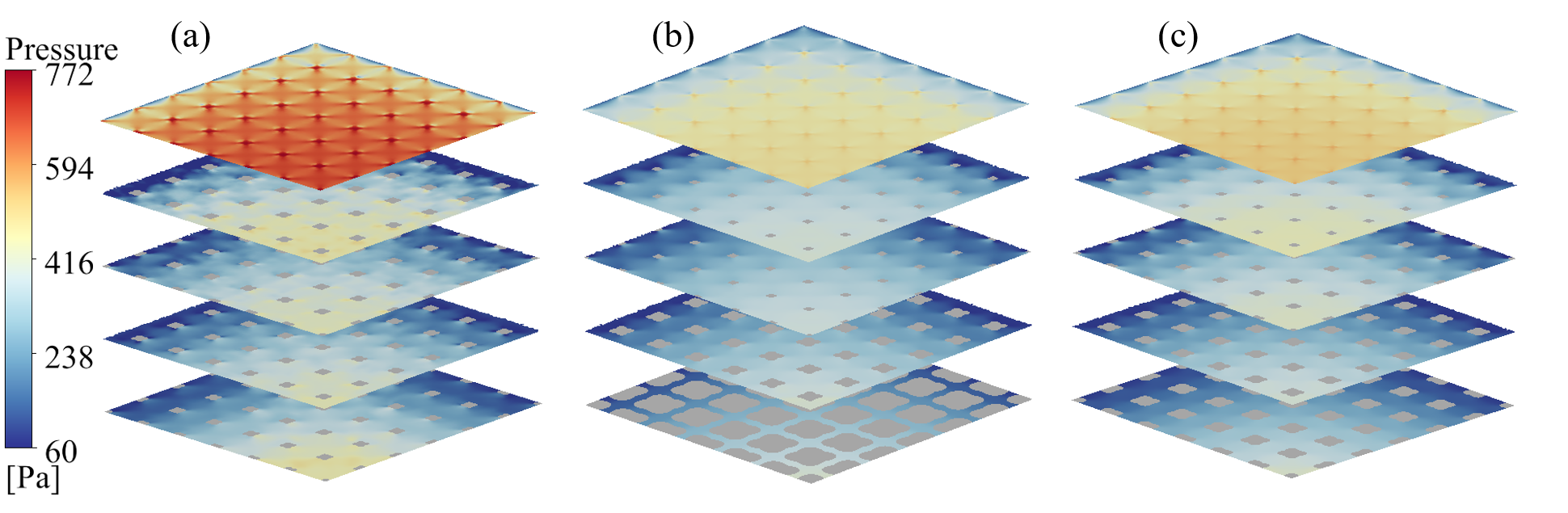}
    \caption{The pressure contours at regular height intervals are shown for (a) the ground lattice, (b) the O1 lattice, and (c) the O2 lattice. The inlet pressure, which represents the average pressure in the topmost layer, is highest in the ground structure. In contrast, the inlet pressures for both the O1 and O2 lattices are significantly lower, supporting the hypothesis. Additionally, the inlet pressure for the O1 lattice is lower than that of the O2 lattice, primarily due to the much thinner struts in its upper layers, although the difference is minimal.}
    \label{fig:fluent_pressure_contours}
\end{figure*}

O1 possesses a significantly larger fluid-contact surface area. While its average fluid velocity is lower (see \Cref{fig:fluent_velocity contours}), this vast increase in area is the primary driver for its high Nusselt number and greater overall convective transfer, far compensating for the velocity-based thermal penalty. O1 also features a larger contact area at the lattice-chip interface. This allows more total heat to be extracted from the chip. However, as shown in \Cref{fig:fluent_temperature_contours}, the average steady-state chip temperature is the highest ($\approx$ \SI{345}{\kelvin}) in the case of the O1 lattice, despite its higher Nusselt number. This poor thermal performance can be attributed to flow stagnation. The thicker struts at the bottom of the O1 lattice impede airflow, leading to poor airflow in the critical region directly above the chip. As a result, the incoming air bypasses this restrictive lower zone, flowing mostly through the upper region, as is evident in \Cref{fig:fluent_velocity contours}. The thicker struts also reduce the chip's exposed surface area to the cold air. This localized flow blockage is the dominant effect, starving the chip surface of cold air and preventing effective heat removal, leading to high temperatures. In contrast, the ground ($\approx \SI{337}{\kelvin}$) and O2 ($\approx \SI{340}{\kelvin}$) lattices allow for better air circulation to the chip, resulting in lower and more preferable temperatures.
The O2 lattice also excels by having a lower pressure drop as compared to the ground structure, as shown in \Cref{fig:fluent_pressure_contours}. This is due to the lower flow resistance facilitated by having thinner struts at the top in the O2 lattice. In summary, as compared to the ground lattice, the O1 lattice excels with a higher Nusselt number and a lower pressure drop, but results in a much higher chip temperature, whereas the O2 lattice excels with an almost identical Nusselt number, a lower pressure drop, with only a slight increase in the average chip temperature. The O2 lattice is the clear choice, as it extracts more heat from the chip than the O1 lattice, despite having a lower Nusselt number, resulting in a lower average chip temperature.
\subsection{Trade-Offs Between Competing Objectives}
\begin{figure}[htbp]
    \centering
    \begin{subfigure}{0.48\textwidth}
        \centering
        \includegraphics[width=\linewidth]{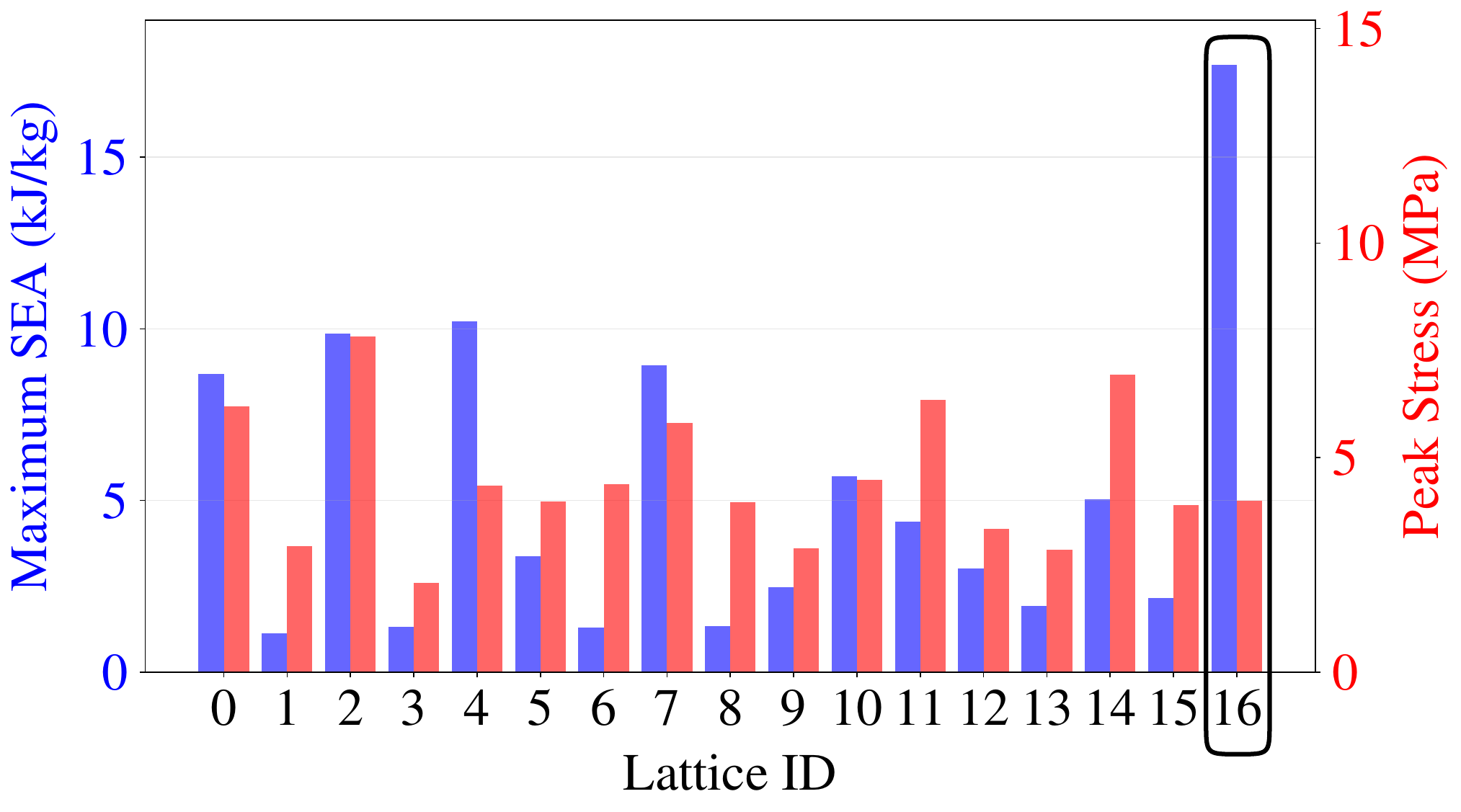}
        \caption{} 
        \label{fig:impact_barplot}
    \end{subfigure}
    \hfill
    \begin{subfigure}{0.48\textwidth}
        \centering
        \includegraphics[width=\linewidth]{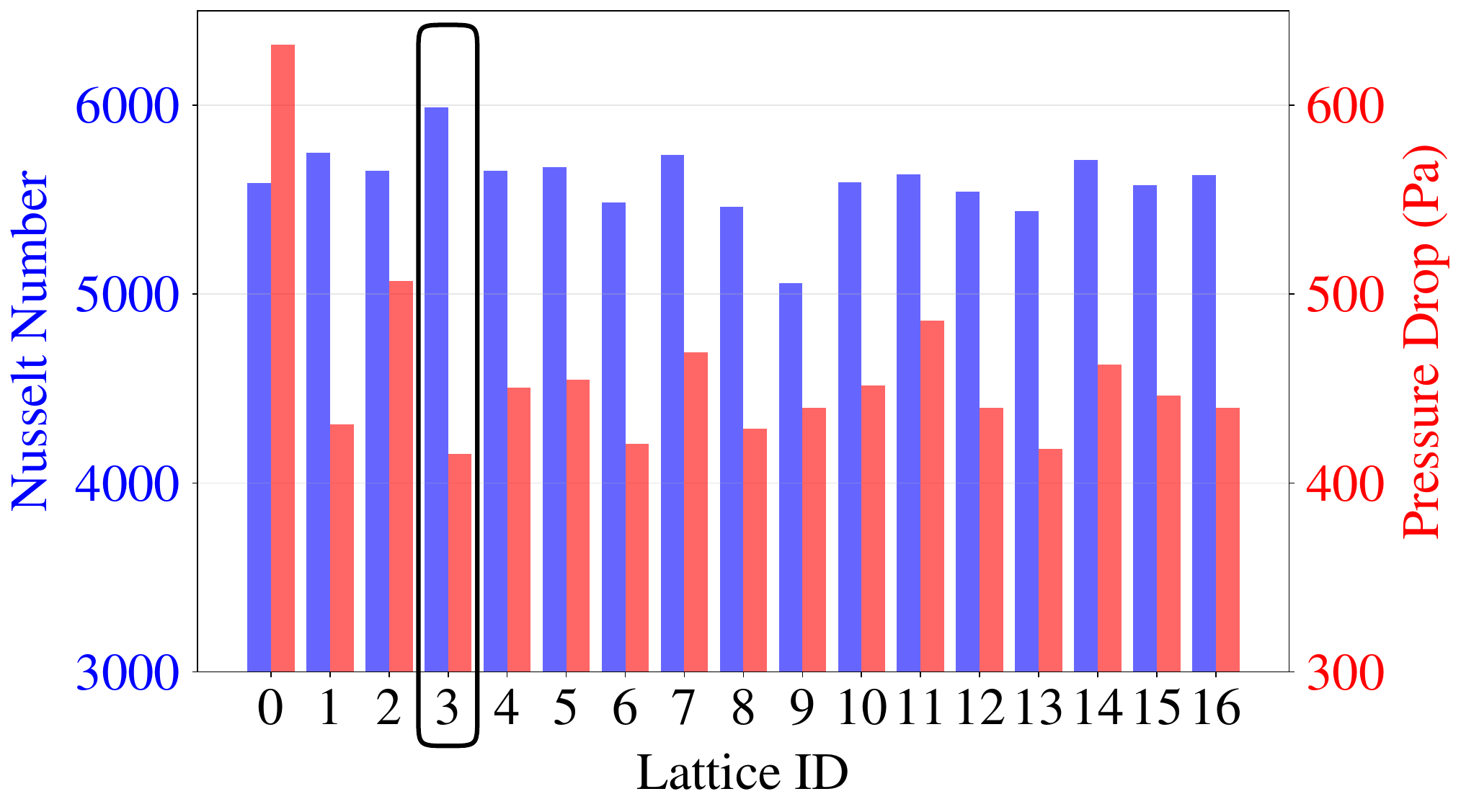}
        \caption{} 
        \label{fig:fluent_barplot}
    \end{subfigure}
    
    \caption{Bar plots showing the (a) impact performance (described by maximum SEA and peak stress) and (b) thermal performance (described by Nusselt number and pressure drop) of each lattice design, including the ground lattice. As highlighted by the black bounding boxes, FG-BCC-16 and lattice FG-BCC-3 have the best impact performance and thermal performance, respectively.}
    \label{fig:performance_comparison}
\end{figure}

\begin{figure}[htbp]
    \centering
    \begin{subfigure}{0.48\textwidth}
        \centering
        \includegraphics[width=\linewidth]{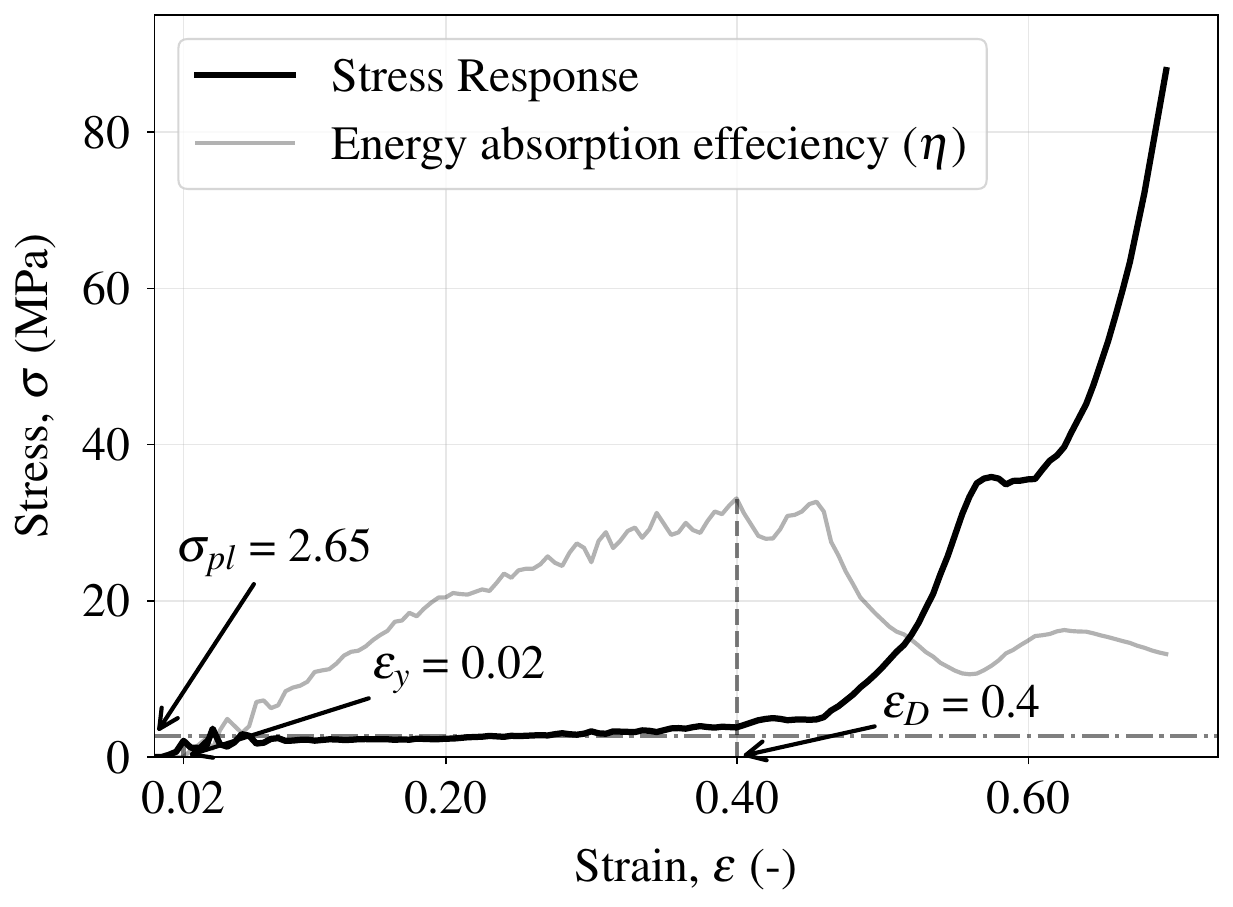}
        \caption{}
        \label{fig:lattice3_stressStrain}
    \end{subfigure}
    \hfill
    \begin{subfigure}{0.48\textwidth}
        \centering
        \includegraphics[width=\linewidth]{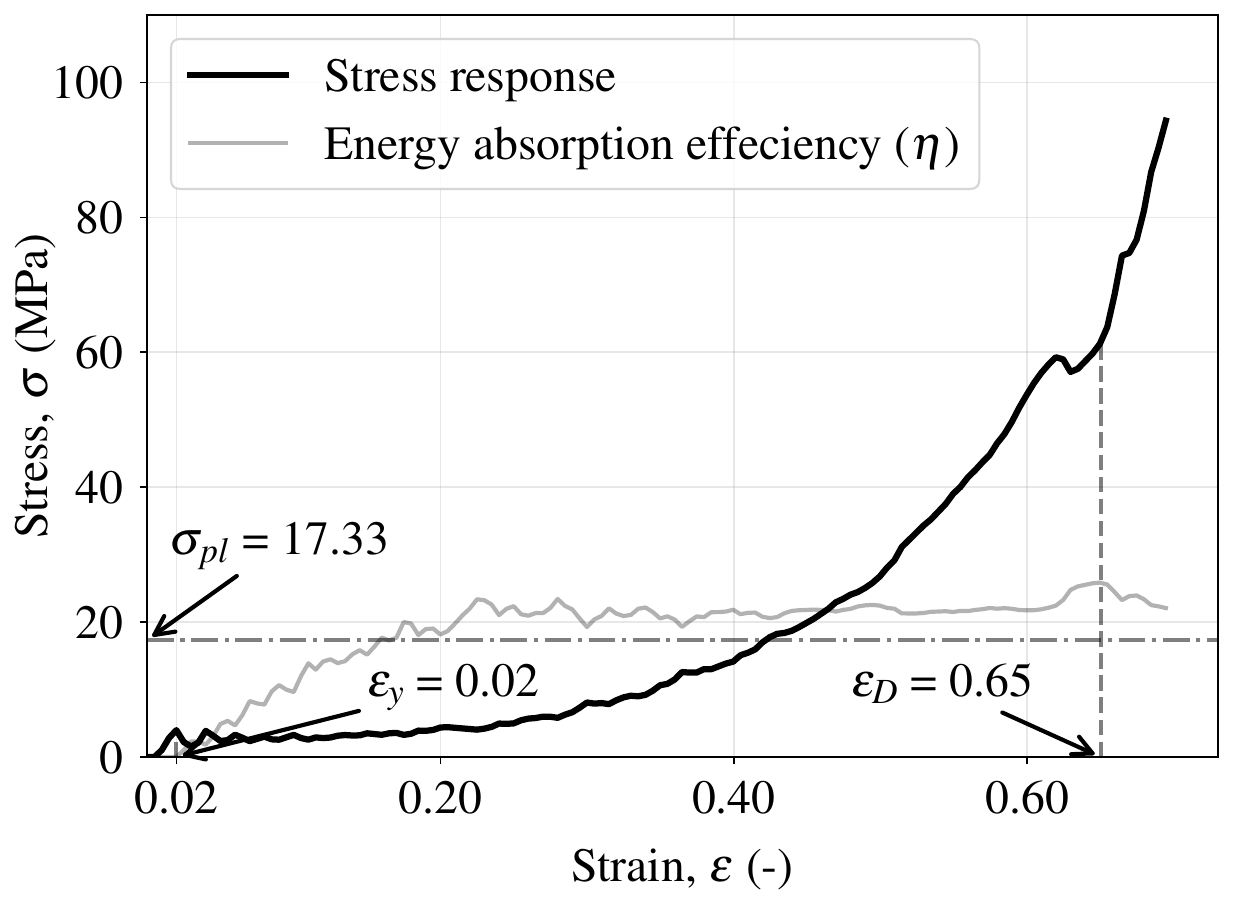}
        \caption{} 
        \label{fig:lattice16_stressStrain}
    \end{subfigure}
    \caption{The stress-strain and energy efficiency profiles of two lattice types: (a) the FG-BCC-3 lattice demonstrates a significant double densification phenomenon attributed to its compound lattice-like behavior, while (b) the FG-BCC-16 lattice exhibits deformation similar to the O2 lattice but has a less pronounced second densification phase.}
    \label{fig:stress_strain_comparison}
\end{figure}

\begin{figure*}[t]
    \centering
    \includegraphics[width=\linewidth]{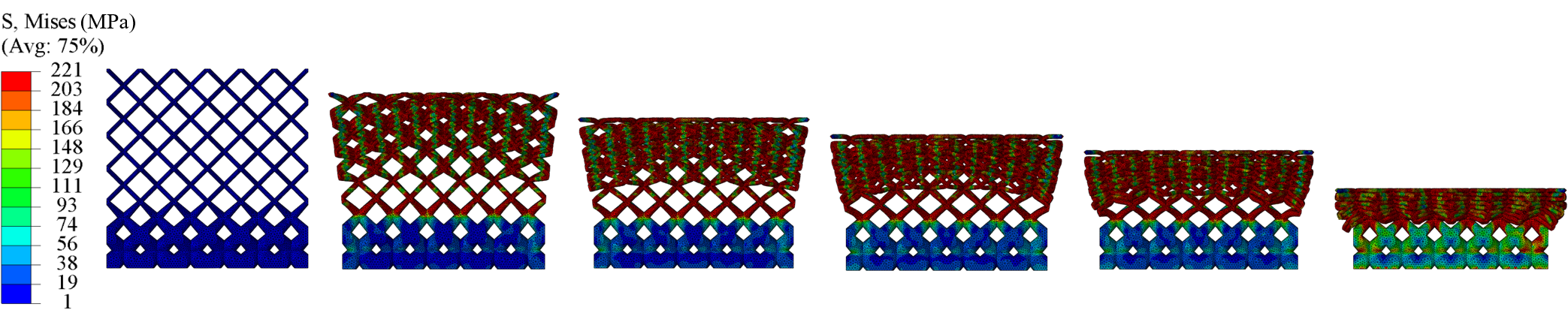}
    \put(-405,0){$\varepsilon=0$}
    \put(-345,0){$\varepsilon=0.125$}
    \put(-273,0){$\varepsilon=0.225$}
    \put(-200,0){$\varepsilon=0.325$}
    \put(-128,0){$\varepsilon=0.400$}
    \put(-52,0){$\varepsilon=0.590$}
    \vspace{2mm}
    \caption{The deformation pattern in the FG-BCC-3 lattice exhibits a compound lattice-like behavior, where the lower half of the lattice begins to deform only after the upper half has been densified. The existence of an almost uniform strut diameter of 0.5 mm for two-thirds of the height of the FG-BCC-3 lattice results in suboptimal performance (in $0 < \varepsilon < 0.6$) as compared to the ground lattice that absorbs more impact energy owing to its larger uniform diameter of 0.78 mm.}
    \label{fig:lattice3_deformation}
\end{figure*}

The O1 and O2 lattices, which are Pareto-optimal designs for different sets of weights, result from carefully balanced trade-offs between impact and thermal performance. A careful examination of the simulation data from the 16 training points is conducted to reveal the geometric features that favor one property over the other. \cref{fig:impact_barplot} and \cref{fig:fluent_barplot} summarize the impact and thermal performance of all 16 training points. As highlighted by the black bounding boxes, the FG-BCC lattices, FG-BCC-16 and FG-BCC-3, exhibit the best performance in impact absorption and thermal dissipation, respectively.
    \begin{table}[t]
    \centering
    \normalsize
    \caption{The design features of lattice-3 and lattice-16}
    \label{tab:design_3_16}
     \setlength{\tabcolsep}{2pt}
    \begin{tabular}{c c c c c}
    \hline
    Lattice ID & $d_0$ (mm) & $d_1$ (mm) & $d_2$ (mm) & $d_3$ (mm) \\
    \hline
    3  & 0.500 & 0.519 & 0.520 & 2.452 \\
    16 & 0.500 & 0.557 & 0.897 & 1.334 \\
    \hline
    \end{tabular}
\end{table}
\Cref{tab:design_3_16} shows that these lattices vary significantly in $d_2$ and $d_3$, whereas $d_1$ is nearly identical, differing by only one decimal place. Moreover, it is to be noted that FG-BCC-3 and FG-BCC-16 have similar diameter values as O1 and O2 lattices, respectively. In the case of FG-BCC-3, the diameter values do not increase significantly until the last diameter ($d_3$), resulting in a steep diameter gradient in the third planar zone closest to the heat source. Due to this, FG-BCC-3's behavior can be thought to be similar to a compound lattice made by placing a lattice of an almost uniform strut diameter of \SI{0.5}{\mm} placed on top of a second lattice whose struts vary in diameter abruptly to a maximum limit (from $d_2$ to $d_3$). This last lattice portion behaves almost like a dense solid, with no ability to undergo plastic deformation. This is exactly what is observed in the FG-BCC-3 impact simulation results. \cref{fig:lattice3_stressStrain} shows the stress-strain profile for FG-BCC-3 under deformation. We observe an almost flat plateau region due to lattice deformation with very slender struts, until the initial densification at around a strain of 0.4. Then it continues to densify to a strain of 0.6, at which point the collapse of the penultimate unit-cell layer begins, as shown in \Cref{fig:lattice3_deformation}, resulting in a slight drop in stress, followed by an increase again.
This shows that FG-BCC-3 behaves as expected, like a compound lattice. Since the slender section of the lattice undergoes greater strain at very low stress, it absorbs minimal plastic energy, even compared to the ground structure, which has thicker (\SI{0.78}{\mm}) but uniform struts than FG-BCC-3. FG-BCC-16, on the other hand, behaves very similarly to the O2 lattice due to the resemblance in diameter values. Hence, as the first couple of layers of FG-BCC-16 are slender, similar to the O2 lattice, the diameter increases steadily, resulting in gradually climbing stress values, as shown in \Cref{fig:lattice16_stressStrain}. Consequently, more plastic energy is absorbed throughout the plastic deformation phase. Hence, although FG-BCC-3 was exceptional in thermal performance, its impact performance is subpar due to its inherent geometry, resulting in a compound-lattice behavior.

On the other hand, examining the design similarity between FG-BCC-16 and the O2 lattice revealed that it had the best impact performance and an overall improved thermal performance, despite the lower Nusselt number.

\subsection{Implications of the PIGO Correlation Analysis}
\label{subsec:pigo_discussion}

The correlation results reveal that PIGOs do not replace raw diameter variables for surrogate modeling. Instead, their importance lies in providing physical interpretability and broader simultaneous coverage across multiple outputs than any single diameter value achieves alone.

\subsubsection{$\nabla\varphi$ is the Most Important Operator}

The porosity gradient $\nabla\varphi$ is the standout operator across the entire analysis, yielding the strongest broad-coverage result in the dataset. No single raw variable replicates this: $d_1$ achieves strong $\sigma_p$ and $\Delta P$ correlations individually but is blind to SEA, while $d_2$ captures SEA well but is essentially uncorrelated with $\Delta P$. $\nabla\varphi$ achieves both simultaneously because it encodes the effective density contrast along the gradation axis using the nonlinear porosity-diameter relationship, rather than a raw diameter endpoint. A steep positive $\nabla\varphi$ simultaneously reduces flow entry resistance and ensures slender struts at the impact face, explaining its dual predictive power. This is the central finding of the PIGO analysis.

\subsubsection{SEA Resists Operator Prediction and $d_2$ Remains the Best Predictor}

No operator surpasses $d_2$ for SEA ($r = 0.721$), nullifying the original PIGO hypothesis. The best operator competitor is $\nabla^2 D$ ($r = -0.684$), which falls below the threshold. SEA accumulates nonlinearly through the entire collapse history and depends 
on the full diameter profile in a way that no scalar can compress without information loss. $d_2$ is therefore retained in the recommended input set on purely empirical grounds.

\subsubsection{Nu is Irreducibly Two-Dimensional}

No variable exceeds $|r| = 0.31$ for Nu. The near-constant $\tau$ values (range: 0.492–0.497) are particularly revealing: the three-zone gradation redistributes mass along the height but barely changes the fraction of surface area in the thermally active mid-zone, because the total surface area itself shifts with gradation in a way that keeps the zonal fraction nearly constant, preventing any scalar from capturing the competing effects of surface area and flow blockage that govern Nu. This is consistent with the nonlinear saddle-like Nu surrogate surface in \Cref{fig:Nu_surrogate}. Prediction of Nu requires either the full unconstrained $(d_1, d_2)$ input space or a field-level descriptor such as a layer-by-layer surface area profile $A_\mathrm{sf}(z)$.

\subsubsection{Recommended Input Set}

The recommended surrogate input set is $[d_2,\nabla\varphi,\Omega,A_\mathrm{sf},J_z]$. Each component earns its place: $d_2$ anchors SEA prediction, $\nabla\varphi$ provides the broadest multi-output coverage, $\Omega$ dominates $\Delta P$ prediction with 
$r = -0.965$, $A_\mathrm{sf}$ captures mesh-level surface geometry relevant to heat transfer, and $J_z$ adds independent mass distribution information that strengthens both $\sigma_p$ and $\Delta P$ predictions. The Nu degeneracy and the SEA dominance of $d_2$ are the two open challenges that future work should address.

\section{Conclusions and Future Work}
By exploring the possibility of optimizing an FG-BCC lattice using surrogate modeling and multi-objective optimization techniques, we have successfully identified two optimal lattice designs, O1 and O2. In comparison to the ground lattice,
\begin{itemize}
    \item the O1 lattice is unsuitable for heat extraction, leading to a significant spike in chip temperature (despite an 8\% increase in Nu). Furthermore, it performs very poorly under impact, showing an 88\% reduction in absorbed energy (SEA). However, it exhibits a lower peak stress ($\sigma_\text{p}$) by 67\% and pressure drop ($\Delta P$) by 35\%;
    \item the O2 lattice excels in energy absorption, showing a 115\% improvement in SEA and a 36\% decrease in $\sigma_\text{p}$. Importantly, it is also effective for heat extraction, maintaining a reasonable chip temperature, and reducing $\Delta P$ by 31\%.
\end{itemize}

Simulations on O1 and O2 validate the optimization framework, with results closely matching the goal-programming predictions. Almost all FG-BCC designs exhibit lower pressure drop than the ground lattice, confirming the initial hypothesis. The O2 lattice is the clear choice for this application, excelling across all four objectives simultaneously.

Extending this study, Pearson correlation analysis shows that raw diameter variables remain competitive and cannot be wholly replaced by PIGOs. The mid-zone diameter remains the strongest SEA predictor, while the porosity gradient stands out among PIGOs by simultaneously predicting both peak stress and pressure drop strongly, which is something no single raw variable achieves. The Nusselt number resists prediction by all variables tested, confirming that heat transfer is irreducibly multidimensional in this configuration. The recommended surrogate input set combines the empirical strengths of the mid-zone diameter, porosity gradient, impingement surface porosity, total interface area, and axial mass moment.

Future work should focus on three directions. First, formal cross-validation comparing surrogate fit quality between $(d_1, d_2)$ and the recommended PIGO set would quantify the improvement concretely. Second, Nu prediction likely requires a field-level input such as a layer-by-layer surface area profile $A_\mathrm{sf}(z)$, which a neural network surrogate could exploit. Third, the operator definitions are topology-agnostic and extend directly 
to FCC, octet, diamond cubic, and TPMS lattices, broadening the applicability of this framework to a wide class of architected materials.

\section{Funding}
No funding was received for conducting this study.

\section{Conflict of Interest}
The authors have no competing interests.

\section{Data Availability Statement}
All data generated or analyzed during this study are included in this article and its supplementary information files.

\appendix 
\setcounter{table}{0}
\renewcommand{\thetable}{A\arabic{table}}
\section{Final Design Points}
\label{app:simulation_values}
\Cref{tab:geometry_info} summarises the list of design points chosen for training the surrogate model.

\begin{table}[htbp] 
\centering
\caption{The final design points, ($d_1$, $d_2$), considered for the training dataset, including the ground structure}
\label{tab:geometry_info}
\begin{tabular}{cccccccc}
\toprule
Lattice ID & $d_0$ (mm) & $d_1$ (mm) & $d_2$ (mm) & $d_3$ (mm) & Porosity (\%) & Mass (g) & Porosity Error (\%) \\
\midrule
0  & 0.780 & 0.780 & 0.780 & 0.780 & 82.950 & 6.250 & --   \\
1  & 0.500 & 0.503 & 0.937 & 1.381 & 81.141 & 7.068 & 2.182 \\
2  & 0.500 & 0.825 & 0.838 & 0.871 & 82.670 & 6.510 & 0.338 \\
3  & 0.500 & 0.519 & 0.520 & 2.452 & 78.010 & 8.260 & 5.955 \\
4  & 0.500 & 0.589 & 0.975 & 1.456 & 81.950 & 6.830 & 1.194 \\
5  & 0.500 & 0.669 & 0.754 & 1.393 & 81.790 & 6.840 & 1.398 \\
6  & 0.500 & 0.513 & 0.694 & 1.956 & 79.730 & 7.620 & 3.218 \\
7  & 0.500 & 0.705 & 0.898 & 0.996 & 82.430 & 6.600 & 0.629 \\
8  & 0.500 & 0.537 & 0.814 & 1.582 & 80.910 & 7.180 & 2.459 \\
9  & 0.500 & 0.616 & 0.640 & 1.817 & 80.490 & 7.330 & 2.966 \\
10 & 0.500 & 0.615 & 0.869 & 1.261 & 81.880 & 6.810 & 1.289 \\
11 & 0.500 & 0.764 & 0.772 & 1.137 & 82.410 & 6.610 & 0.651 \\
12 & 0.500 & 0.593 & 0.726 & 1.652 & 80.910 & 7.170 & 2.459 \\
13 & 0.500 & 0.524 & 0.597 & 2.199 & 78.940 & 7.920 & 4.834 \\
14 & 0.500 & 0.690 & 0.828 & 1.177 & 82.220 & 6.680 & 0.880 \\
15 & 0.500 & 0.603 & 0.797 & 1.454 & 81.470 & 6.970 & 1.784 \\
16 & 0.500 & 0.557 & 0.897 & 1.334 & 81.520 & 6.940 & 1.723 \\
\bottomrule
\end{tabular}
\end{table}

\section{Simulation Results}
The simulation results obtained for the four different parameters across the 16 FG-BCC lattices are given in \Cref{tab:full_simulation_results}.

\begin{table}[htbp]
\centering
\caption{Simulation results for the 16 design points}
\label{tab:full_simulation_results}
\begin{tabular}{cccccccc}
\toprule
Lattice ID & SEA (kJ/kg) & $\sigma_\text{p}$ (MPa) & $\sigma_\text{pl}$ (MPa) & Nu & $\Delta P$ (Pa) & $C_\text{f}$ & $\beta$ \\
\midrule
1 & 13.83 & 4.34 & 14.69 & 5747.20 & 431.19 & 3.13 & 3929.43 \\
2 & 9.87 & 7.83 & 11.17 & 5653.63 & 506.89 & 3.68 & 3662.57 \\
3 & 1.31 & 2.08 & 2.65 & 5988.10 & 415.60 & 3.02 & 4144.70 \\
4 & 10.22 & 4.35 & 12.91 & 5653.92 & 450.45 & 3.27 & 3809.75 \\
5 & 3.37 & 3.97 & 6.39 & 5673.07 & 454.64 & 3.30 & 3810.88 \\
6 & 1.29 & 4.38 & 3.03 & 5485.81 & 420.69 & 3.05 & 3781.67 \\
7 & 8.93 & 5.81 & 11.85 & 5737.16 & 469.24 & 3.40 & 3813.54 \\
8 & 1.35 & 3.96 & 3.31 & 5462.65 & 428.86 & 3.11 & 3741.63 \\
9 & 2.48 & 2.88 & 4.66 & 5059.10 & 439.96 & 3.19 & 3435.83 \\
10 & 5.70 & 4.48 & 8.78 & 5592.04 & 451.69 & 3.28 & 3764.61 \\
11 & 4.38 & 6.35 & 8.26 & 5632.34 & 485.98 & 3.53 & 3700.37 \\
12 & 3.02 & 3.33 & 5.07 & 5543.46 & 439.99 & 3.19 & 3764.69 \\
13 & 1.94 & 2.85 & 3.39 & 5440.39 & 418.16 & 3.03 & 3757.90 \\
14 & 5.04 & 6.93 & 7.74 & 5711.82 & 462.53 & 3.36 & 3814.97 \\
15 & 2.17 & 3.89 & 4.59 & 5577.95 & 446.15 & 3.24 & 3770.60 \\
16 & 17.70 & 3.99 & 17.33 & 5630.88 & 439.90 & 3.19 & 3824.32 \\
\bottomrule
\end{tabular}
\end{table}

\section{Calculated Values of PIGOs}
\label{app:pigo_tables}

\Cref{tab:pigo_values} reports the computed value of each Physics-Informed Geometric 
Operator for all 17 FG-BCC lattice designs.

\begin{table}[htbp]
\centering
\caption{Computed PIGO values for all 17 FG-BCC lattice designs.}
\label{tab:pigo_values}
\begin{tabular}{ccccccccccc}
\toprule
ID & $\nabla^2 D$ & $\nabla\varphi$ & $\Omega$ & $A_\mathrm{sf}$ & $\kappa_\mathrm{int}$ & $\eta_I$ & $\chi$ & $J_z$ & $\tau$ \\
\midrule
0 (Ground) & 0.000 & 0.000 & 1.000 & 11029.02 & 5.694 & 1.000 & 0.169 & 0.258 & 0.497 \\
1  & 0.003 & 0.062 & 1.058 & 10713.68 & 8.551 & 0.908 & 0.188 & 0.228 & 0.495 \\
2  & 0.005 & 0.029 & 1.016 & 10977.13 & 6.968 & 0.940 & 0.172 & 0.249 & 0.496 \\
3  & 0.483 & 0.108 & 1.125 &  9993.19 & 6.846 & 0.908 & 0.219 & 0.201 & 0.493 \\
4  & 0.024 & 0.066 & 1.046 & 10841.59 & 9.557 & 0.914 & 0.179 & 0.237 & 0.496 \\
5  & 0.139 & 0.063 & 1.045 & 10850.85 & 9.304 & 0.975 & 0.181 & 0.237 & 0.496 \\
6  & 0.270 & 0.090 & 1.089 & 10425.37 & 8.561 & 0.921 & 0.202 & 0.214 & 0.496 \\
7  & $-$0.024 & 0.037 & 1.021 & 10932.52 & 9.840 & 0.938 & 0.175 & 0.238 & 0.497 \\
8  & 0.123 & 0.073 & 1.059 & 10664.07 & 8.916 & 0.929 & 0.190 & 0.226 & 0.495 \\
9  & 0.288 & 0.084 & 1.076 & 10624.44 & 9.481 & 0.970 & 0.194 & 0.225 & 0.496 \\
10 & 0.035 & 0.055 & 1.036 & 10840.60 & 9.849 & 0.942 & 0.180 & 0.236 & 0.496 \\
11 & 0.089 & 0.047 & 1.025 & 10953.99 & 8.888 & 0.975 & 0.175 & 0.246 & 0.496 \\
12 & 0.198 & 0.076 & 1.063 & 10690.52 & 9.626 & 0.957 & 0.190 & 0.228 & 0.496 \\
13 & 0.382 & 0.100 & 1.109 & 10248.24 & 8.634 & 0.921 & 0.210 & 0.210 & 0.494 \\
14 & 0.053 & 0.049 & 1.028 & 10908.91 & 9.493 & 0.960 & 0.177 & 0.241 & 0.496 \\
15 & 0.116 & 0.066 & 1.043 & 10779.25 & 9.856 & 0.953 & 0.184 & 0.234 & 0.496 \\
16 & 0.024 & 0.059 & 1.035 & 10770.52 & 9.295 & 0.926 & 0.184 & 0.231 & 0.496 \\
\bottomrule
\end{tabular}
\end{table}

\FloatBarrier
\bibliography{bibliography_with_doi}

\end{document}



\title[Article Title]{Hybrid Surrogate-Based Multi-Objective Optimization of Graded BCC Lattices}

\author[1]{\fnm{Jaswanth} \sur{Vellore Gurudev}}\email{vgjaswanth@gmail.com}

\author*[2]{\fnm{Ratna Kumar} \sur{Annabattula}}\email{ratna@iitm.ac.in}

\affil[1]{\orgdiv{Department of Metallurgical and Materials Engineering}, \orgname{Indian Institute of Technology Madras}, \orgaddress{\city{Chennai}, \postcode{600036}, \state{Tamil Nadu}, \country{India}}}

\affil*[2]{\orgdiv{Mechanics of Materials Laboratory, Department of Mechanical Engineering}, \orgname{Indian Institute of Technology Madras}, \orgaddress{\city{Chennai}, \postcode{600036}, \state{Tamil Nadu}, \country{India}}}


\maketitle

\section{Derivation for the equivalent diameter, $D$}
\label{app:appendix1}
\begin{figure*}[t]
    \centering
    \includegraphics[width=\linewidth]{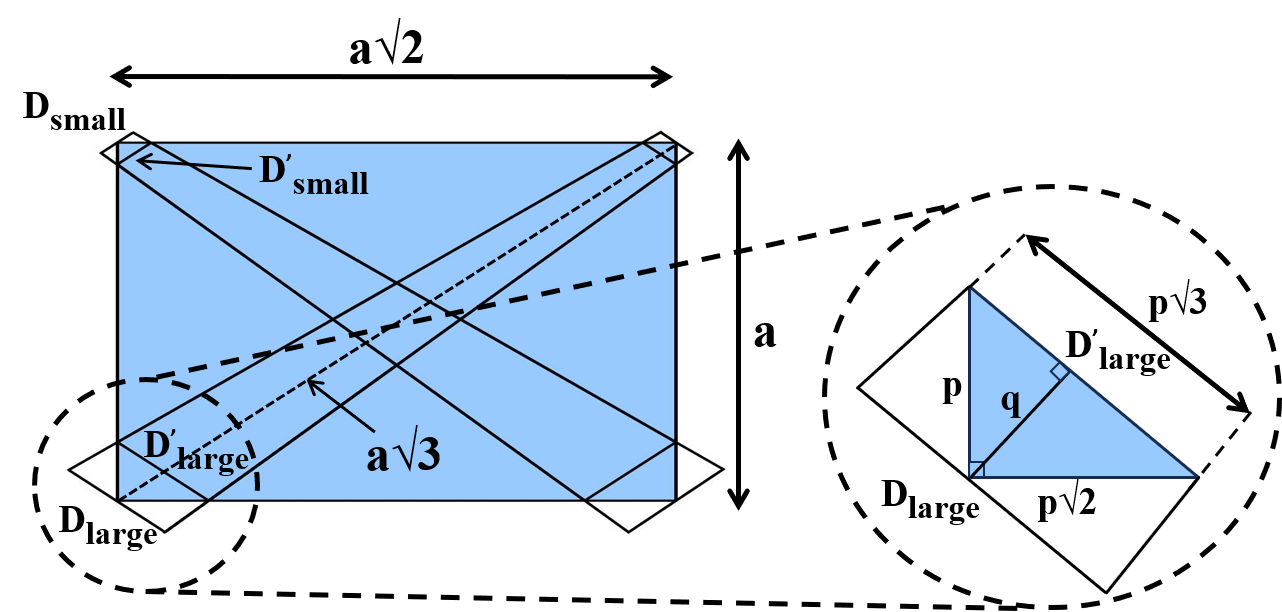}
    \caption{The zoomed-in representation is shown to understand the calculation of the parallel interior diameters}
    \label{fig:zoomed_diameter}
\end{figure*}
Leveraging the linear variation in diameter over the diagonal of length $ a\sqrt {3}$, from \Cref{fig:zoomed_diameter}, we can write,

\begin{equation}
\begin{aligned}
    D_\text{large}^{'} &= p\sqrt{3}
    = D_\text{large} + \frac{p\sqrt{2}}{3a}\big(D_\text{small} - D_\text{large}\big), \\
    &p \bigg[\sqrt{3} - \frac{\sqrt{2}\big(D_\text{small} - D_\text{large}\big)}{3a} \bigg] = D_\text{large}, \\
    p &= \frac{D_\text{large}}{\sqrt{3} - \frac{\sqrt{2}}{3a}\big(D_\text{small} - D_\text{large}\big)}, \\
    D_\text{large}^{'} &= p\sqrt{3}
    = \frac{D_\text{large}}{1 - \frac{\sqrt{2}}{3\sqrt{3} a}\big(D_\text{small} - D_\text{large}\big)}.
\end{aligned}
\end{equation}

Similarly, we get, 

\begin{equation}
    D_\text{small}^{'} = \frac{D_\text{small}}{1 + \frac{\sqrt{2}}{3\sqrt{3} a}\big(D_\text{small} - D_\text{large}\big)}.
\end{equation}

We find the equivalent diameter, $D$, by taking average of $D_\text{small}^{'}$ and $D_\text{large}^{'}$:

\begin{equation}
\begin{split}
    D = \frac{D_\text{small}^{'}+D_\text{large}^{'}}{2}  
    = \frac{1}{D_\text{small} - D_\text{large}}\Bigg(\frac{D_\text{small}}{2 + \frac{2 \sqrt{2}}{3 \sqrt{3} a}} + \frac{D_\text{large}}{2 - \frac{2 \sqrt{2}}{3 \sqrt{3} a}}\Bigg)
\end{split}
\end{equation}

\section{Derivation for the area porosity}
\label{app:appendix2}
\begin{figure}[htbp]
    \centering
    \includegraphics[width=0.8\linewidth]{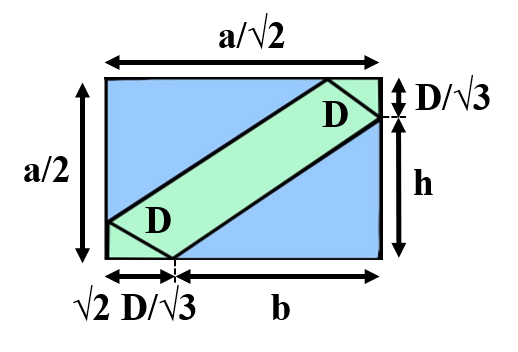}
    \caption{One quarter of the cross-section containing an equivalent uniform lattice with struts of diameter, D, is shown, which occupies the same area as the functionally graded lattice. The area porosity is now simply calculated by dividing the area of the unoccupied blue region by the total area}
    \label{fig:sub-rectangle}
\end{figure}

The area porosity, which is the ratio of the unoccupied area (the blue region) to the total area of the rectangle, is given by
\begin{equation}
    \text{Area Porosity} = \frac{2 \times \frac{1}{2}  b  h}{\frac{a}{\sqrt{2}} \times \frac{a}{2}}
\end{equation}
where, $b$ and $h$ are the dimensions of the blue triangle as shown in \Cref{fig:sub-rectangle}, whose lengths are defined as 
\begin{equation}
    b = \frac{a}{\sqrt{2}} - \frac{\sqrt{2}D}{\sqrt{3}} \;\;\;\;\; \text{and} \;\;\;\;\; h = \frac{a}{2} - \frac{D}{\sqrt{3}}
\end{equation}

Simplifying the expression, the area porosity is finally given by
\begin{equation}
\begin{split}
    \text{Area Porosity} =  \frac{2\times 0.5 \times\bigg(\frac{a}{\sqrt{2}} - \frac{\sqrt{2}D}{\sqrt{3}}\bigg)\bigg( \frac{a}{2} - \frac{D}{\sqrt{3}}\bigg)}{\frac{a}{\sqrt{2}} \times \frac{a}{2}} = \bigg(1 - \frac{2D}{a \sqrt{3}}\bigg)^2.
\end{split}
\end{equation}

\section{Sampling data points using Latin Hypercube Sampling (LHS)}
With $d_\text{max}$ = \SI{2.58}{mm} as our upper bound for all $d_1$, $d_2$ and $d_3$, combined with $d_0$ $<$ $d_1$ $<$ $d_2$ $<$ $d_3$ to serve our hypothesis of having thicker struts at the base, we naturally have $d_1 \in [d_0,d_2]$, $d_2\in [d_1, d_3]$ and $d_3\in [d_2, 2.58]$.

\begin{figure}[H]
    \centering
    \begin{subfigure}{\linewidth}
        \centering
        \includegraphics[width=0.7\linewidth]{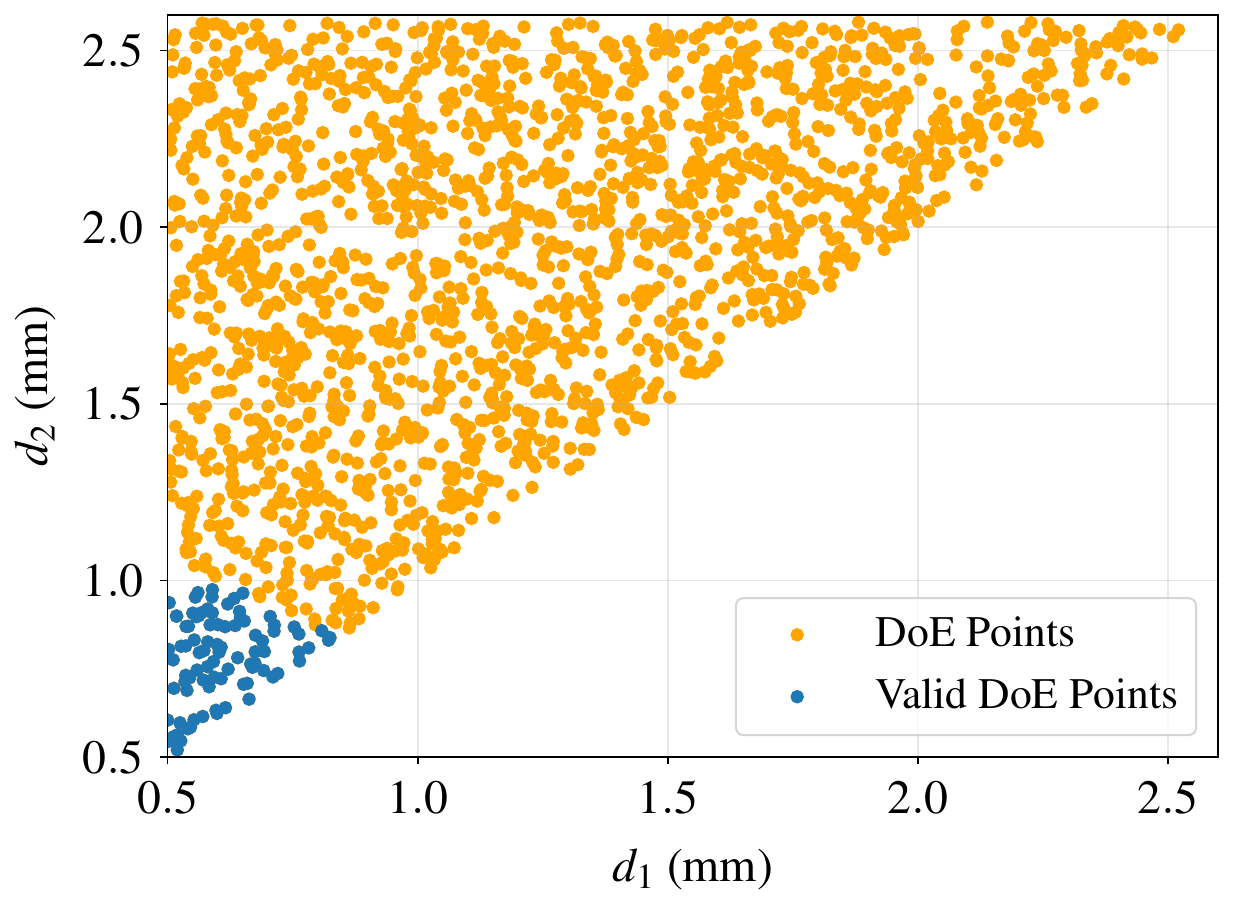}
    \caption{}
    \label{fig:d1_lessthan_d2}
    \end{subfigure}

    \begin{subfigure}{\linewidth}
        \centering
        \includegraphics[width=0.7\linewidth]{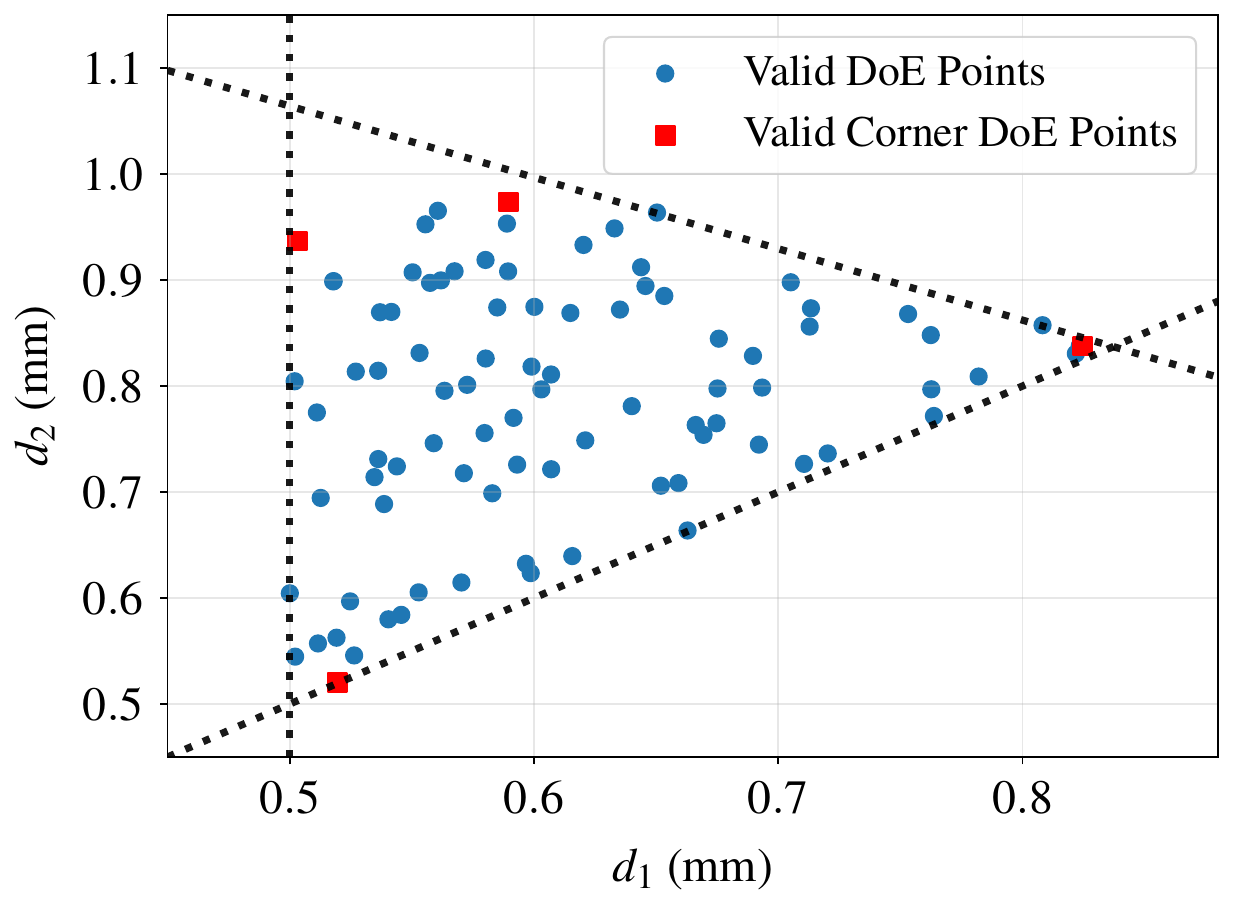}
    \caption{}
    \label{fig:mid_Selection_points}
    \end{subfigure}

    \caption{(a) The set of all ($d_1$, $d_2$) DoE points (with $d_1 < d_2$) generated using LHS (orange) and the set of valid DoE points satisfying the area porosity constraint (blue), and (b) the zoomed-in plot of all valid points showing the naturally arising feasible region bounded by the dotted lines and red corner points}
    \label{fig:doe_selection_process}
\end{figure}

To choose 16 points within our bounds as defined above and subject to density constraints, we first generate a large number of ($d_1$, $d_2$) points within the range [0.5, 2.58] as shown in \Cref{fig:d1_lessthan_d2}. This is achieved using Latin Hypercube Sampling (LHS) to generate uniformly distributed points. Once this is done, we take each and every ($d_1$, $d_2$) point, and then compute $d_3$ for $\rho$ = 0.17 (the corresponding area porosity = 0.399) along with $d_0$ = \SI{0.5}{\mm}. If there exists a suitable $d_3$ in the range [$d_2$, 2.58], then the corresponding ($d_1$, $d_2$) point is further considered for the next round of filtering. Once the set of all ($d_1$, $d_2$) points that satisfy the constraints is generated, we now select four points representing the four corners of our newly obtained feasible region, which is now an embodiment of all the constraints we have defined. 

\begin{figure}[H]
    \centering
    \includegraphics[width=0.7\linewidth]{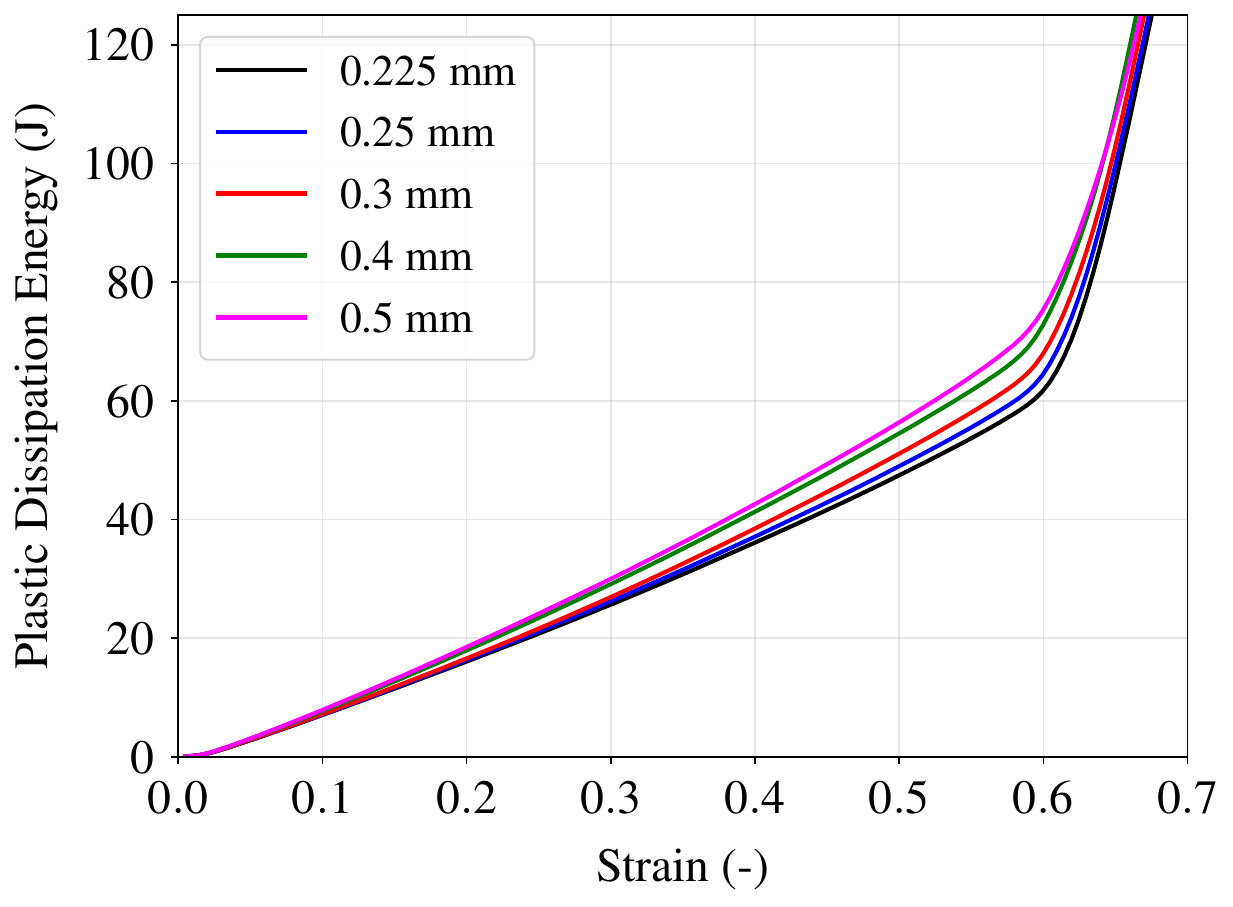}
    \caption{The final set of 16 evenly spaced ($d_1$, $d_2$) points (shown in green) in our feasible region represents our training geometries}
    \label{fig:final_DoE_points}
\end{figure}

\Cref{fig:mid_Selection_points} displays the points that meet the relative density constraint, with the feasible region outlined and the four corner points marked in red. The remaining 12 points are chosen using the Greedy Poisson Disk (GPD) method (which utilizes the concept of minimum distance to choose uniformly spaced points in our feasible region bounded by the corner points) and highlighted in green along with the four red corner points as shown in \Cref{fig:final_DoE_points}. 
\section{Governing equations for SST k-omega formulation}
The SST (Shear Stress Transport) k-omega model combines the k-omega and k-epsilon turbulence models using a blending function. The analytical expressions for the continuity equations for mass conservation and transport equations for momentum conservation (along with their blending function) are as follows:\\
\noindent 1. Continuity equation:
\begin{equation}
    \label{eq:continuity_eqn}
    \ \nabla\cdot\rho\,\textbf{u} = 0
\end{equation}
\noindent 2. Transport equations:\\
The kinetic energy ($k$) - transport equation is:
\begin{equation}
    \label{eq:k-transport}
    \begin{split}
        \frac{\partial (\rho k)}{\partial t} + \frac{\partial (\rho u_j k)}{\partial x_j} &= P_\text{k} - \beta \rho \omega k 
         + \frac{\partial}{\partial x_j} \left[ \left( \mu + \frac{\mu_\text{t}}{\sigma_\text{k}}  \right) \frac{\partial k}{\partial x_j} \right] 
    \end{split}
\end{equation}
where $k =$ \SI{1}{\meter\squared\per\second\squared} is the turbulent kinetic energy, $\rho$ is the density of the fluid, $u_j$ is the mean fluid velocity, $\omega =$ \SI{1}{\second^{-1}} is the specific dissipation rate, $\beta = 0.09$ is a model constant, $\mu$ and $\mu_\text{t}$ are bulk and turbulent viscosity ($\mu_\text{t} / \mu = 10$), respectively, $\sigma_\text{k} = 0.9$ is the turbulent Prandtl number for $k$ and $P_\text{k}$ is the turbulent kinetic energy production term.\\
\\
The specific dissipation rate ($\omega$) - transport equation is:
\begin{equation}
    \label{eq:w-transport}
    \begin{split}
        \frac{\partial{(\rho \omega)}}{\partial{t}} + \frac{\partial{(\rho u_j \omega)}}{\partial{x_j}} &= \frac{\partial}{\partial{x_j}}\Bigg[\left(\mu + \frac{\mu_\text{t}}{\sigma_{\omega}}\right) \frac{\partial{\omega}}{\partial{x_j}}\Bigg] + \frac{\gamma}{\nu_\text{t}}P_\text{k} - \beta \rho \omega^2 + 2(1-F)\frac{\rho \sigma_{\omega}^2}{\omega}\frac{\partial k}{\partial x_j}\frac{\partial{\omega}}{\partial x_j}
    \end{split}
\end{equation}
where $\sigma_{\omega} = 1.0$ is the turbulent Prandtl number for $\omega$, $\gamma$ is the production coefficient, and $\nu_\text{t} = \mu_\text{t}/\rho$ is the turbulent dynamic viscosity. We use a blending function, $(1-F)$, to switch from the k-epsilon model when away from a wall (free stream region) to the k-omega model when close to a wall. When $F=1$, the final term of \Cref{eq:w-transport}, called the cross diffusion term, becomes zero, and hence serves as the transport equation for $\omega$,  like the k-omega model. When moving away from the wall, i.e., $0\leq F<1$, the cross diffusion term remains, and mimics the behaviour of the $\epsilon$-equation by introducing a coupling between k and $\omega$, similar to the way $\epsilon$ behaves in the k-epsilon model. Due to the steady-state assumption, the time derivative term in the transport equations becomes zero.
\section{Mesh convergence study for the CFD model} 
Patch-independent tetrahedral elements were used to mesh the whole CFD model. Upon meshing, it was found that the skewness was at most 0.84 and the lowest orthogonal quality was 0.15. Typically, having more cells helps obtain an accurate result, albeit at the cost of increased computational time. For studying mesh convergence, we have used the volume-averaged chip temperature and inlet pressure as convergence criteria for the forced convection simulation, with an inlet velocity of \SI{1}{\meter\per\second}. Since the lowest dimension in our design is \SI{0.5}{\mm} at the inlet, we use at least 2 cells across the strut's cross-section for accuracy. Adaptive sizing was used to gradually increase the element size from \SI{0.25}{\mm} in regions of higher geometric complexity to \SI{2}{\mm} in regions of lower geometric complexity. Using two cells across the inlet strut's cross-section yielded approximately 4.84 million cells, which met the minimum required number. The number of cells was slowly increased by decreasing the minimum cell size to study convergence. As a result, we see that beyond 5.5 million elements, the variables of interest do not increase drastically as shown in \Cref{tab:fluent_mesh_convergence}. Hence, we proceed with using 5.5 million cells for the forced convection simulations.
\begin{table}[htbp]
    \centering
    \normalsize
    \caption{Mesh convergence results for forced convection simulation.}
    \label{tab:fluent_mesh_convergence}
    \begin{tabular}{ccccc}
        \hline
        Number of cells & Avg. Chip & \% & Inlet Pressure & \% \\ 
        (millions) & Temp. (K) & change & (Pa) & change \\ \hline
        4.84 & 337.766 & --    & 4.4284 & --    \\
        5.50 & 337.341 & 0.130 & 4.4482 & 0.450 \\
        7.00 & 337.328 & 0.004 & 4.4505 & 0.052 \\ \hline
    \end{tabular}
\end{table}
\section{Mesh convergence for the FEA model}
\begin{figure}[H]
    \centering
    \includegraphics[width=0.7\linewidth]{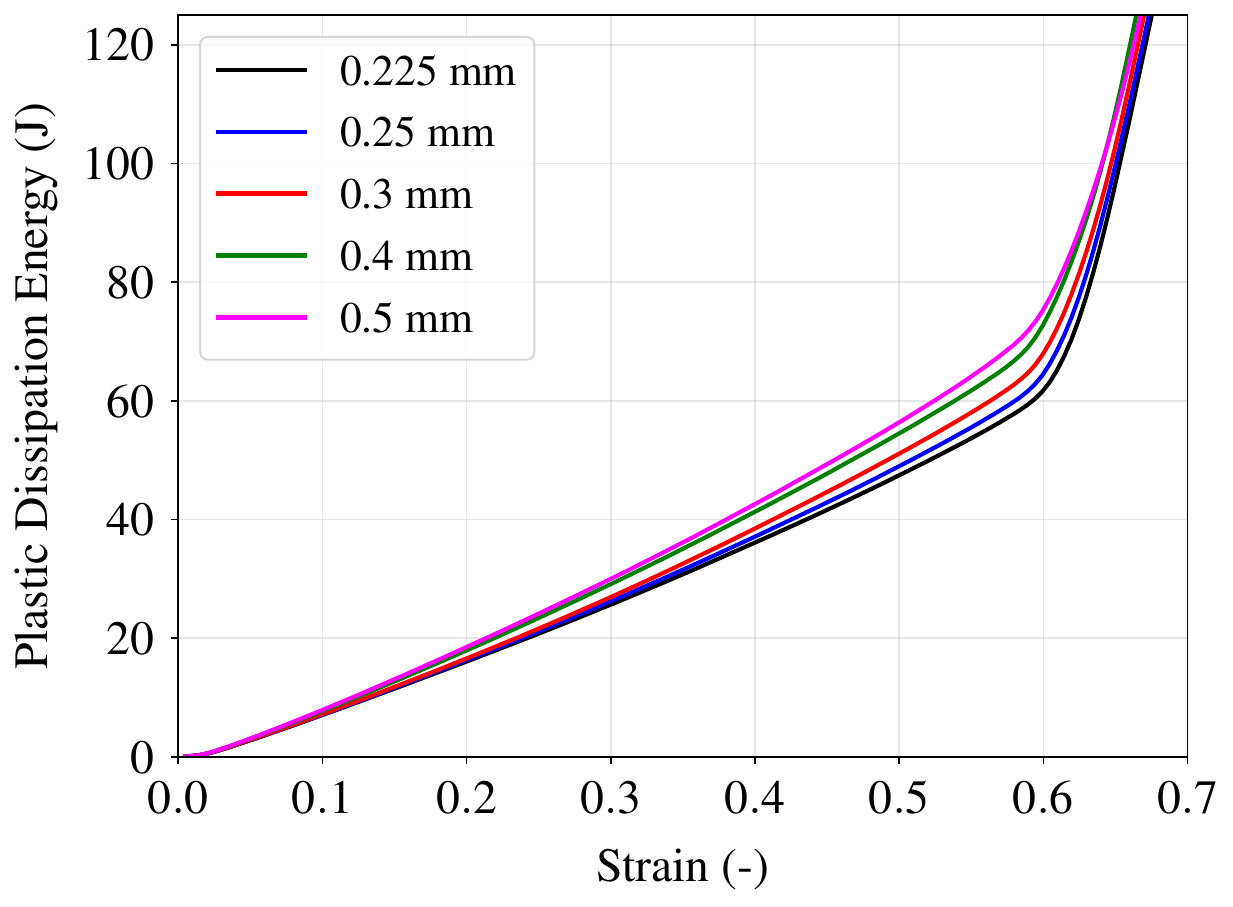}
    \caption{Mesh convergence results of impact simulations on the ground BCC lattice showing negligible improvement below an element size of \SI{0.25}{\mm}}
    \label{fig:Abaqus_meshConvergence}
\end{figure}

Impact simulations were performed for the ground FG-BCC lattice meshed with element sizes of 0.5 mm, 0.4 mm, 0.3 mm, 0.25 mm, and 0.225 mm. The total plastic dissipation energy of all these simulations is plotted to visualize the improvement achieved with finer meshes. As shown in \Cref{fig:Abaqus_meshConvergence}, minimal improvement in accuracy is observed beyond a mesh size of 0.25 mm; therefore, we use this size for the discretization.

